\documentclass[journal,twoside,web, letterpaper]{ieeecolor}
\usepackage{tac}
\usepackage{cite}[sort,compress]
\usepackage{amsmath,amssymb,amsfonts}
\usepackage{mathtools, mathrsfs}
\usepackage{euscript}
\usepackage[pdftex]{graphicx}
\usepackage{blindtext}
\usepackage{datetime}
\usepackage{xspace}
\usepackage{xcolor}
\usepackage{subcaption}
\let\labelindent\relax
\usepackage{enumitem}
\usepackage{comment}
\usepackage[normalem]{ulem}

\DeclareMathAlphabet{\mathpzc}{OT1}{pzc}{m}{it}

\definecolor{mblue}{rgb}{0,0.4470,0.7410}
\definecolor{morange}{rgb}{0.8500,0.3250,0.0980}
\definecolor{myellow}{rgb}{0.9290,0.6940,0.1250}
\definecolor{mpurple}{rgb}{0.4940,0.1840,0.5560}
\definecolor{mgreen}{rgb}{0.4660,0.6740,0.1880}
\definecolor{mcyan}{rgb}{0.3010,0.7450,0.9330}
\definecolor{mred}{rgb}{0.6350,0.0780,0.1840}
\definecolor{mgreenblue}{rgb}{0.0,1.0,0.5}
\definecolor{parulablue}{rgb}{0.2431,0.1490,0.6588}
\definecolor{parulalblue}{RGB}{39,151,235}
\definecolor{parulagreen}{RGB}{129,204,89}
\definecolor{parulayellow}{RGB}{249,251,21}

\definecolor{cblue}{rgb}{0,0.9,1}
\definecolor{corange}{rgb}{1,0.7,0}
\definecolor{olive}{rgb}{0.5,0.5,0}

\usepackage{amsthm}

\theoremstyle{definition}
\newtheorem{defn}{Definition}
\newtheorem{exmp}{Example}
\theoremstyle{plain}
\newtheorem{theorem}{Theorem}
\newtheorem{lemma}{Lemma}
\newtheorem{prop}{Proposition}

\newtheorem{assumption}{Assumption}
\theoremstyle{remark}
\newtheorem{rmrk}{Remark}

\newenvironment{definition}{\begin{defn}}{\hfill$\square$\end{defn}}
\newenvironment{remark}{\begin{rmrk}}{\hfill$\square$\end{rmrk}}

\newcommand{\tss}[1]{\textsuperscript{#1}}
\newcommand{\lpvcore}{\textsc{LPVcore}\xspace}
\newcommand{\matlab}{\textsc{Matlab}\xspace}
\newcounter{ass}

\newcommand{\mc}[1]{{\mathcal{#1}}}
\newcommand{\mf}[1]{{\mathfrak{#1}}}
\newcommand{\mr}[1]{{\mathrm{#1}}}
\newcommand{\mb}[1]{{\mathbb{#1}}}
\newcommand{\ms}[1]{{\mathscr{#1}}}
\newcommand{\mt}[1]{{\mathtt{#1}}}
\newcommand{\msf}[1]{{\mathsf{#1}}}

\newcommand{\meu}[1]{{\EuScript{#1}}}
\newcommand{\mpzc}[1]{{\mathpzc{#1}}}

\newcommand{\possemidef}{\succeq}

\newcommand{\unaryminus}{\scalebox{0.65}[1]{\ensuremath{\,-}}}

\newcommand{\diag}{\mr{diag}}

\DeclareMathOperator{\sinc}{sinc}

\DeclareMathOperator{\proj}{Proj}
\newcommand{\kron}{\otimes} %
\newcommand{\bkron}{\circledcirc}
\newcommand{\svdots}{\raisebox{0pt}{$\scalebox{.75}{\vdots}$}}
\newcommand{\sddots}{\raisebox{0pt}{$\scalebox{.75}{$\ddots$}$}}

\newcommand{\vect}[1]{\mpzc{#1}}

\newcommand{\dnx}{n_\mr{x}}
\newcommand{\dny}{n_\mr{y}}
\newcommand{\dnu}{n_\mr{u}}
\newcommand{\dnp}{n_\mr{p}}
\newcommand{\dna}{n_\mr{a}}
\newcommand{\dnb}{n_\mr{b}}
\newcommand{\dnw}{n_\mr{w}}

\newcommand{\dnr}{n_\mr{r}}

\newcommand{\N}{\mb{N}}

\newcommand{\R}{\mb{R}}

\newcommand{\Bfint}[2]{\left.\mf{B}_{#1}\right|_{#2}}

\newcommand{\dataset}{\mc{D}_{N}}

\newdateformat{monthyeardate}{\monthname[\THEMONTH], \THEYEAR}
\newdateformat{yeardate}{\THEYEAR}

\renewcommand{\comment}[1]{}

\begin{document}
\title{Data-driven Dissipativity Analysis of Linear Parameter-Varying Systems}
\author{%
Chris Verhoek, %
Julian Berberich,
Sofie Haesaert, %
Frank Allg{\"o}wer, and
Roland T{\'o}th %
\thanks{This work has received funding from from the European Research Council (ERC) (grant agreement nr. 714663) and has also been supported by the European Union within the framework of the National Laboratory for Autonomous Systems (RRF-2.3.1-21-2022-00002).
The work has been also funded by Deutsche Forschungsgemeinschaft (DFG, German Research Foundation) under Germany's Excellence Strategy - EXC 2075 - 390740016 and under grant 468094890. 
		We acknowledge the support by the Stuttgart Center for Simulation Science (SimTech).
}
\thanks{C. Verhoek, R. T\'oth and S. Haesaert are with the Control Systems Group, Eindhoven University of Technology, The Netherlands. J. Berberich and F. Allg\"ower are with the University of Stuttgart, Institute for Systems Theory and Automatic Control, Germany. R. T\'oth is also with the Institute for Computer Science and Control, Hungary. Email addresses: \textbraceleft \texttt{c.verhoek}, \texttt{r.toth}, \texttt{s.haesaert}\textbraceright \texttt{@tue.nl}, and \textbraceleft \texttt{julian.berberich}, \texttt{frank.allgower}\textbraceright \texttt{@ist.uni-stuttgart.de}. Corresponding author: Chris Verhoek.}
}

\maketitle
\begin{abstract}
	We derive direct data-driven dissipativity analysis methods for Linear Parameter-Varying (LPV) systems using a single sequence of input-scheduling-output data. By means of constructing a semi-definite program subject to linear matrix inequality constraints based on this \emph{data-dictionary}, direct data-driven verification of $(Q,S,R)$-type of dissipativity properties of the data-generating LPV system is achieved. Multiple implementation methods are proposed to achieve efficient computational properties and to even exploit structural information on the scheduling, e.g., rate bounds. The effectiveness and trade-offs of the proposed methodologies are shown in simulation studies of academic and physically realistic examples.

\end{abstract}


\section{Introduction}\label{sec:introduction}
\IEEEPARstart{D}{irect} data-driven methods are attractive to analyze system behavior and generate stabilizing controllers from data without the need of identifying a mathematical description of the system. A cornerstone {of} direct data-driven analysis and control for discrete-time  \emph{linear time-invariant} (LTI) systems is the so-called \emph{Fundamental Lemma} by Willems et al. \cite{WillemsRapisardaMarkovskyMoor2005}. This result uses LTI behavioral system theory \cite{PoldermanWillems1997} to obtain a characterization of the system behavior based on a single sequence of \emph{input-output} (IO) data. Based on the Fundamental Lemma, numerous results have been developed for LTI systems, e.g., on data-driven simulation~\cite{MarkovskyRapisarda2008}, system analysis~\cite{romer2019one,koch2021provably,vanWaarde2022_matrixS}, and (predictive) control \cite{markovsky2007linear, dePersisTesi2020, coulson2019data,BerberichKohlerMullerAllgower2021DPC_guarantees}, many of which also guarantee robustness in the presence of noise.
Some extensions towards the \emph{nonlinear} system domain have been made as well, e.g., for Hammerstein and Wiener systems \cite{BerberichAllgower2020}, \emph{linear time-varying} (LTV) \cite{nortmann2020data} systems, flat nonlinear systems \cite{alsalti2021data}, nonlinear autoregressive exogenous (NARX) systems~\cite{mishra2021narx}, and bilinear systems~\cite{yuan2021data}. However, these results impose heavy restrictions on the {underlying} system as they leverage model transformations and linearizations. A particularly interesting extension of Willems' Fundamental Lemma is the one towards \emph{linear parameter-varying} (LPV) systems \cite{VerhoekAbbasTothHaesaert2021, VerhoekTothHaesaertKoch2021, VerhoekTothHaesaert2023, Verhoek2022_DDLPVstatefb, Verhoek2022_DDLPVstatefb_experiment}.

LPV systems are linear systems, where the model parameters, describing the linear signal relation, are dependent on a time-varying variable, referred to as the \emph{scheduling variable}. The latter variable is used to express nonlinearities, time-variations, or exogenous effects. The main difference with respect to LTV systems is that the scheduling variable is \emph{not} known a priori; it is only assumed that it is measurable and allowed to vary in a given set. The LPV framework has been shown to be able to capture a relatively large subset of nonlinear systems in terms of LPV surrogate models \cite{Toth2010} and can therefore be considered as a promising extension towards data-driven analysis and control for nonlinear systems.

In this paper, we focus on direct data-driven dissipativity analysis of LPV systems. The dissipativity property of a system simultaneously gives guarantees on the stability and performance characteristics of the system. This makes dissipativity a \emph{fundamental building block} in analysis and control of dynamical systems \cite{SchererWeiland2021}, especially because it is an essential tool in performance-based controller synthesis. While theory on direct data-driven dissipativity verification for LTI systems is rather established \cite{romer2019one, koch2020verifying, koch2021provably, vanWaarde2022_matrixS}, including handling of noisy data and robust verification, the literature on dissipativity analysis for LPV systems (and nonlinear systems) is mainly focussed on \emph{model-based} approaches. To perform analysis in a data-based setting, it is often required to have an \emph{indirect} data-driven approach, i.e., first identifying the system, followed by a model-based analysis on the estimated model. There are some direct data-driven dissipativity analysis approaches for nonlinear systems available based on a set-membership argument \cite{TangDaoutidis2021, martin2021approximation, martin2021data, martin2021dissipativity, MartinAllgower2022}. 
However, these methods are not based on the behavioral framework and often require restrictive assumptions, %
e.g., output measurements can only be handled by the construction of an extended state that must be controllable, or it is assumed that full state measurements are available from the system. 
Therefore, having a tool based on the behavioral LPV framework that allows for direct data-driven analysis of the dissipativity property of an LPV system using only scheduling and IO data, significantly contributes towards a data-driven framework with guarantees for LPV systems, and possibly a direct data-driven framework for nonlinear systems (see, e.g., \cite{Verhoek2022_DDLPVstatefb_experiment,10384139} for a promising result).

Our contributions in this paper are the following:
\begin{enumerate}[label={C\arabic*:}, align=left, ref={C\arabic*}, leftmargin=*]
	\item Characterization of finite-horizon dissipativity for LPV systems by means of a fully data-driven representation; \label{C:char}
	\item {Computational methods with different levels of conservatism and computational tractability} 
	for direct dissipativity analysis of a class of LPV systems using \emph{only} input-scheduling-output data; \label{C:comp}
	\item Extensive analysis of the methods in simulation studies, {including the influence of measurement noise on the proposed approaches}. \label{C:simu}
\end{enumerate}

\smallskip

In this work, we consider LPV systems that can be {described}
 by an LPV-IO representation with \emph{shifted-affine} scheduling dependency. This realization form is highly attractive in practice, cf.~\cite{kwiatkowski2006automated, sloth2011robust, HoffmannWerner2014, deLange2022lpv, Toth2011_SSrealizationTCST}, %
as it allows the application of powerful prediction error minimization approaches, but at the same time, due to a direct state-space realization with static-affine scheduling dependency, it also allows the use of well-established convex analysis and control synthesis approaches on the model estimates. Constituting to Contribution~\ref{C:comp}, the presented direct data-driven dissipativity analysis techniques can be efficiently solved as a \emph{semi-definite program} (SDP), subject to a finite set of \emph{linear matrix inequality} (LMI) constraints, similar to  model-based analysis approaches, {but without requiring an analytic LPV model of the} system.

In the remainder of this work, we first present the problem setting in Section~\ref{sec:problemstatement}, followed by an introduction of the data-driven LPV representation in Section~\ref{sec:ddlpvrep}. Based on this, %
we {introduce a data-driven characterization of} dissipativity in
Section~\ref{sec:datadrivendissipativity}, corresponding {to} Contribution~\ref{C:char}. We derive computable analysis approaches
in Section~\ref{sec:computational}, providing Contribution~\ref{C:comp}. Finally, we {show applicability and analyze properties of the proposed methods  on simulation examples} in Section~\ref{sec:examples}, {corresponding to Contribution~\ref{C:simu},} and draw our conclusions on the presented results in Section \ref{sec:conclusions}. %

\subsection{Notation}
{The set of positive integers is denoted {as} $\mathbb{N}$, while $\mathbb{R}$ denotes the set of real numbers.}
The $p$-norm of a vector $x_k\in\mathbb{R}^{n_\mathrm{x}}$  is denoted by $\lVert x_k\rVert_p$, while  {$x_{k,j}$} stands for the $j$\tss{th} element of $x_k$. The identity matrix of size $n$ is given by $I_n$. For a matrix $A\in\mathbb{R}^{n\times m}$,  $A^\dagger$ stands for its Moore-Penrose inverse, while $A^\perp \in\mathbb{R}^{q\times n}$ corresponds to a  matrix of {row rank $q$}, where the rows form the basis of the left kernel of $A$ , i.e., $A^\perp A=0$. Furthermore, $A\kron B$ is the Kronecker product of two matrices $A$ and $B$. Given square matrices $A_1,\dots,A_n$, $\diag_{i=1}^n(A_i)$ gives a block-diagonal matrix with blocks $A_i$. For vector spaces $\meu{A}$ and $\meu{B}$,  $\proj_\meu{B}(\meu{A})$ denotes the projection of $\meu{A}$ onto $\meu{B}$. 
For a sequence $w_{[1,N]}: [1,N] \rightarrow \mathbb{R}^{n_\mathrm{w}}$, we denote
the Hankel matrix of depth $L$ associated with {it} %
as
\begin{align*}
\mc{H}_L(w_{[1,N]}) = \begin{bmatrix} 
    w_1 & w_2 & \cdots & w_{N-L+1} \\
    w_2 & w_3 & \cdots & w_{N-L+2} \\
    \vdots & \vdots & \ddots & \vdots \\
    w_{L} & w_{L+1} & \cdots & w_{N}
\end{bmatrix}.
\end{align*}

\section{Problem setting}
\label{sec:problemstatement}
As mentioned in Section~\ref{sec:introduction}, an attractive approach to address identification and potentially control design in the LPV framework is to {describe} systems with discrete-time \emph{shifted-affine} LPV-IO representations {\cite{Toth2011_SSrealizationTCST, HoffmannWerner2014}}.
Hence, we consider that the underlying system is {represented} by %
\begin{subequations}\label{eq:sys}
\begin{align}\label{eq:lpv_shi_aff}
y_k+\sum_{i=1}^{\dna}a_i(p_{k-i})y_{k-i}=\sum_{i=0}^{\dnb}b_i(p_{k-i})u_{k-i},
\end{align}
with input $u_k\in\mathbb{R}^{\dnu}$, output $y_k\in\mathbb{R}^{\dny}$ and scheduling signal $p_k\in\mathbb{P}\subseteq\mathbb{R}^{\dnp}$, $\dna\geq1$, $\dnb\ge0$, and $k\in\N$ being the discrete time, where $\mb{P}$ is a convex and compact set {called} the \emph{scheduling space}. Let $n_\mathrm{r}=\max(\dna,\dnb)$. {The initial condition for \eqref{eq:lpv_shi_aff} is defined as $\{y_k\}_{k=1-\dna}^{0}$, $\{p_k\}_{k=1-n_\mathrm{r}}^{0}$ and $\{u_k\}_{k=1-\dnb}^{0}$, collected into the vector $w_0\in \mathbb{W}_0=\mathbb{R}^{\dna \dny}\times \mathbb{P}^{\dnr} \times \mathbb{R}^{\dnb \dnu}$.} {Due to the embedding principle, the scheduling trajectories are often restricted to a set $\ms{P}$, e.g., $\mb{P}^\N$, which describes the admissible {behavior of the signal $p_k$.} %
We will give more examples of definitions for $\ms{P}$ in Section~\ref{sec:computational}.} The functions $a_i$ and $b_i$  define the shifted-affine character of the representation, i.e., $a_i$ and $b_i$ are affine functions of the time-shifted values of the scheduling $p_k$:
\begin{align}\label{eq:staticaffinefunc}
a_i(p_{k-i})=\sum_{j=0}^{\dnp}a_{i,j}p_{k-i,j}, \quad b_i(p_{k-i})=\sum_{j=0}^{\dnp}b_{i,j}p_{k-i,j},
\end{align}
\end{subequations}
where $p_{k,0}=1$ for all $k$. It is assumed that $q^{n_\mathrm{r}}(1+\sum_{i=1}^{\dna}a_i(p_{k-i})q^{-i})$ and $q^{n_\mathrm{r}}\sum_{i=0}^{\dnb}b_i(p_{k-i})q^{-i}$  are coprime {and compose a full row rank matrix polynomial in the shift-operator $q$}, implying that the LPV-IO representation \eqref{eq:sys} is minimal and structurally controllable and observable \cite{Toth2010}.

The LPV system represented by \eqref{eq:sys} is characterized in terms of all possible solutions of \eqref{eq:sys}:
\begin{multline}\label{eq:behav:sys}
\mf{B}:=\big\{(u,p,y)\in\left(\R^{\dnu}\times \mb{P}\times \R^{\dny}\right)^\N \mid p\in\ms{P} \text{ and} \\ \exists w_0 \in \mathbb{W}_0 \text{ s.t. \eqref{eq:sys} holds}\ \forall k \in \mathbb{N} \big\},
\end{multline}
which we call the behavior of \eqref{eq:sys}.
Similarly, we can define the projected IO behavior $\mf{B}_{\mr{IO}}$, as the set of admissible IO trajectories, i.e., $\mf{B}_{\mr{IO}}=\pi_{u,y}\mf{B}$, {where $\pi_{u,y}$ stands for  $\proj_{(\R^{\dnu}\times \R^{\dny})^\N} (\mf{B})$}. Also note that $\pi_{p}\mf{B}=\ms{P}$. {Moreover, we define $\dnx$ as the minimal required state-dimension among all possible LPV state-space realizations that can represent $\mf{B}$}, which we will refer to as the \emph{system order}. The \emph{lag} associated with $\mf{B}$ is the minimal $n_\mathrm{r}$ required to represent $\mf{B}$ in terms of~\eqref{eq:sys}. {For {\emph{multiple-input multiple-output} (MIMO)} systems, $\dnx\geq\dnr$, while for \emph{single-input single output} (SISO) systems $\dnx=\dnr$.}
{Finally, we introduce $\mf{B}_p$, which denotes the set that contains all IO trajectories compatible with some scheduling signal $p\in\ms{P}$,}
\begin{equation}\label{eq:behav:p}
    \mf{B}_p=\{(u,y)\in\left(\R^{\dnu}\times \R^{\dny}\right)^\N \mid  (u,p,y)\in\mf{B} \}.
\end{equation}

Analyzing the \emph{behavior} allows us to study the system trajectories independent of the underlying representation. Therefore, the behavioral framework is a helpful tool in assessing stability and performance properties in the analysis and control of systems. As highlighted in Section~\ref{sec:introduction}, an {attractive approach} for simultaneously evaluating \emph{stability} and \emph{performance} is the concept of dissipativity, which is defined in \cite{HillMoylan1980} from an IO perspective in discrete-time as follows.
\begin{definition}[Dissipativity,  \cite{HillMoylan1980}]\label{def:dissipativity-hillmoylan}
    A system with input $u:\N\to\R^{\dnu}$, output $y:\N\to\R^{\dny}$ and behavior $\mf{B}$ is dissipative with respect to the supply rate $\Pi\in\mb{R}^{(\dnu+\dny)\times(\dnu+\dny)}$ if %
    \begin{equation}\label{eq:dissipationinequality}
    \sum_{k={1}}^{N} \begin{bmatrix} u_k \\ y_k \end{bmatrix}^{\!\top}\!\! \Pi \begin{bmatrix} u_k \\ y_k \end{bmatrix}\geq0 \quad \text{for any } N\in\mathbb{N},
    \end{equation}
    and $\forall (u,y)\in\mf{B}_{\mr{IO}}$ with $w_0=0$ (zero initial condition).
\end{definition}
In this paper, we consider the case of QSR-dissipativity, i.e., the supply rate $\Pi$ is partitioned as
\begin{equation}\label{eq:supplyrate}
\Pi=\begin{bmatrix} Q & S \\ S^\top & R \end{bmatrix}, 
\end{equation}
with $Q=Q^\top \in\mb{R}^{\dnu\times\dnu}$, $S\in\mb{R}^{\dnu\times\dny}$, and $R=R^\top \in\mb{R}^{\dny\times\dny}$. This form for the supply rate is commonly used to analyze {system properties}, e.g., passivity %
with $(Q,S,R)=(0,I,0)$ or {computation of an $\ell_2$-gain upper bound} $\gamma$  %
with $(Q,S,R)=(\gamma^2 I,0,\unaryminus I)$.

There are multiple {model-based} techniques to verify stability and performance via \eqref{eq:dissipationinequality} for LTI \cite{Willems1972, HillMoylan1980}, LPV~{\cite{Wu1995, KoelewijnToth2021}~and nonlinear \cite{Schaft2017_L2book, VerhoekKoelewijnToth2021_incremental}} systems. %
{However, when we have} only measured data available from %
{the system,} dissipativity analysis via Definition~\ref{def:dissipativity-hillmoylan} becomes a {difficult} problem, as we can only consider {a few} \emph{finite-time} trajectories. {For this reason, we introduce a finite-time notion of dissipativity, $L$-dissipativity, which allows for analysis with system trajectories of finite length.} Hence, for a given signal $w\in\left(\R^{\dnw}\right)^\N$ and a compact set $\mb{T}\subset\N$, we denote by $w_\mb{T}$ %
the truncation of~$w$ to the time interval $\mb{T}$. Similarly, the restricted behavioral set $\Bfint{}{\mb{T}}$  contains the truncation of all the trajectories in $\mf{B}$ to the time interval $\mb{T}$. In this paper, we consider the following definition of $L$-dissipativity, originally proposed in \cite{maupong2017lyapunov}.
\begin{definition}[$L$-dissipativity, \cite{maupong2017lyapunov}]\label{def:L_diss}
A system with input $u:\N\to\R^{\dnu}$, output $y:\N\to\R^{\dny}$ and behavior $\mf{B}$ is $L$-dissipative w.r.t. the supply rate \eqref{eq:supplyrate} when
\begin{align}\label{eq:def_L_diss}
\sum_{k=1}^{r}\begin{bmatrix}u_k\\y_k\end{bmatrix}^{\!\top} \!\!
\Pi\begin{bmatrix}u_k\\y_k\end{bmatrix}\geq0 \quad \text{for any } r\in[1,L]
\end{align}
and for all truncated system trajectories $(u_{[1,L]}, y_{[1,L]})\in\Bfint{\mr{IO}}{[1,L]}$ with $w_0=0$ (zero initial condition).
\end{definition}
Our main results on data-driven dissipativity analysis for LPV systems are built upon this definition. We are now ready to formulate the problem that we solve in this paper.
\subsubsection*{Problem statement} Consider a data-generating system that can be represented with \eqref{eq:sys}. Given a length $N$ data set $\dataset:= \{u_k,p_k,y_k\}_{k=1}^N$ (data-dictionary) sampled from the data-generating system, i.e., $(u_{[1,N]},p_{[1,N]},y_{[1,N]})\in\Bfint{}{[1,N]}$. For an $1\leq L<N$, how {can we} determine that \eqref{eq:sys} is $L$-dissipative w.r.t. the supply rate \eqref{eq:supplyrate} {using only $\dataset$}?

{If we can solve this problem, then} we can analyze %
performance properties of  {LPV} systems in a data-driven setting, \emph{without} the need of knowing their {dynamic} model \eqref{eq:sys}.

\section{Data-driven LPV representations}
\label{sec:ddlpvrep}
To properly analyze properties of a system, some form of \emph{representation} of the underlying behavior is required. However, one may lose information in a modeling process by estimating such a representation from data. In this section, we focus on a purely data-based representation of \eqref{eq:sys} that is used to analyze dissipativity of the system directly from data. %
We can obtain such a representation by characterizing the behavior of \eqref{eq:sys}, restricted to a finite time-interval $[1,\,L]$, i.e., $\Bfint{}{[1,\,L]}$.

In \cite{VerhoekTothHaesaertKoch2021}, a data-driven representation for general LPV systems has been derived based on behavioral LPV system theory \cite{Toth11_LPVBehav}. 
Using this fundamental result, a simplified data-driven representation has been given in \cite{VerhoekAbbasTothHaesaert2021} for LPV systems that have a representation in terms of \eqref{eq:sys}. This data-driven representation under shifted-affine dependency is obtained by structuring the data in $\dataset$ in terms of \eqref{eq:sys}, such that for valid solution trajectories of the system, i.e., $(\bar{u}_{[1,L]},\bar{p}_{[1,L]},\bar{y}_{[1,L]}) \in \Bfint{}{[1,L]}$, there exists a $g\in\mathbb{R}^{N-L+1}$ that satisfies 
\begin{equation}\label{eq:thm_FL}
	\begin{bmatrix}
		\mc{H}_L\left(u_{[1,N]}\right) \\ 
		\mc{H}_L\left(y_{[1,N]}\right) \\ 
		\mc{H}_L\left(u_{[1,N]}^{\mt{p}}\right) - \bar{\mc{P}}^{\dnu}\mc{H}_L\left(u_{[1,N]}\right) \\ 
		\mc{H}_L\left(y_{[1,N]}^{\mt{p}}\right) - \bar{\mc{P}}^{\dny}\mc{H}_L\left(y_{[1,N]}\right)
	\end{bmatrix}g = \begin{bmatrix}
		\bar{\vect{u}}_{L} \\
		\bar{\vect{y}}_L \\
		0 \\
		0
	\end{bmatrix}.
\end{equation}
Here, for a sequence $\bar{u}_{[1,N]}: [1,N] \rightarrow \mathbb{R}^{n_\mathrm{u}}$,  its vectorization $\mr{vec}\left(\bar{u}_{[1,N]}\right)$ is defined as 
\(
	\bar{\vect{u}}_{\,[1,N]} = \begin{bmatrix} \bar{u}_1^\top & \cdots & \bar{u}_{N}^\top \end{bmatrix}^\top
\)
and we often write $\bar{\vect{u}}_{N}$ for $\bar{\vect{u}}_{\,[1,N]}$. Furthermore, $u_{[1,N]}^{\mt{p}}$ denotes the sequence $\{p_k\kron u_k\}_{k=1}^N$. The same notation is defined for $y$ respectively. Finally in \eqref{eq:thm_FL}, $\bar{\mc{P}}^{n}:=\bar{p}_{[1,L]}\bkron I_{n}$, where $\bkron$ is the block-diagonal Kronecker operator, i.e., for a sequence $\bar{p}_{[1,L]}$ we have $\bar{p}_{[1,L]}\bkron I_n:=\mathrm{blkdiag}_{i=0}^{L}(\bar{p}_i\kron I_n)$. See \cite{VerhoekAbbasTothHaesaert2021} for a detailed derivation of \eqref{eq:thm_FL}.

To represent \emph{any} length $L$ trajectory of \eqref{eq:sys}, the data set $\dataset$ must be sufficiently rich. {The LPV Fundamental Lemma for LPV systems of the form~\eqref{eq:sys}, i.e., the Extended LPV Fundamental Lemma from~\cite{VerhoekTothHaesaert2023} provides a condition to characterize all possible solutions in $\Bfint{\bar{p}}{[1,L]}$ for a given $\bar{p}_{[1,L]}$.
\begin{prop}[Extended LPV Fundamental Lemma]\label{prop:FLeasy}
    Given a data set $\dataset$ from an LPV system represented by~\eqref{eq:sys}. For a $\bar{p}_{[1,L]}\in\ms{P}_{[1,L]}$, define the spaces
    \begingroup\allowdisplaybreaks
	\begin{subequations}\label{eq:nullrowdef}
	\begin{align}
		\meu{N}_{\bar{p}}:= & \mr{nullspace}\left\lbrace\begin{pmatrix}
		\mc{H}_L(u^{\mt{p}}_N) - \bar{\mc{P}}^{\dnu}\mc{H}_L(u_N) \\
		\mc{H}_L(y^{\mt{p}}_N) - \bar{\mc{P}}^{\dny}\mc{H}_L(y_N)
		\end{pmatrix} \right\rbrace, \\ 
		\meu{S} := & \mr{rowspace}\left\lbrace\begin{pmatrix}
		\mc{H}_L(u_N) \\
		\mc{H}_L(y_N)
		\end{pmatrix} \right\rbrace.
	\end{align}
	\end{subequations}
	\endgroup
	For ${L}\ge\dnr$, the data set $\dataset$ satisfies
    \begin{equation} \label{eq:thm:dim}
        \mr{dim}\big\{\proj_{\meu{N}_{\bar{p}}}(\meu{S})\big\} = \dnx + \dnu L,
    \end{equation}
    for all $\bar{p}_{[1,L]}\in\ms{P}_{[1,L]}$, if and only if
    \begin{equation} \label{eq:thm:beh}
        \proj_{\meu{N}_{\bar{p}}}(\meu{S}) = \Bfint{\bar{p}}{[1,L]}
    \end{equation}
    for all $\bar{p}_{[1,L]}\in\ms{P}_{[1,L]}$. This is equivalent to the existence of a vector $g\in\mb{R}^{N-L+1}$ for any $(\bar{u}_{[1,L]},\bar{p}_{[1,L]},\bar{y}_{[1,L]})\in\Bfint{}{[1,L]}$ such that~\eqref{eq:thm_FL} holds.
\end{prop}
\begin{proof}
    See~\cite{VerhoekTothHaesaert2023}.
\end{proof}
We want to emphasize here that this result allows to fully describe an LPV system, whose representation is in the form of \eqref{eq:sys}, using \emph{only} a data set that satisfies condition~\eqref{eq:thm:dim}. This condition in-fact reminiscent of the data set coming from the LPV system being \emph{persistently exciting}\footnote{As condition~\eqref{eq:thm:dim} also contains the output {from} $\dataset$, similar to~\cite{MarkovskyDorfler2021}, {we can} refer to~\eqref{eq:thm:dim} as a \emph{generalized} LPV PE condition.} (PE) of a certain degree. As in the technical definition for PE in~\cite{VerhoekTothHaesaertKoch2021}, we see that the condition is, next to the input, also dependent on the scheduling. This introduces an input-design problem for the generation of a PE $\dataset$, which is {similar to the problem of PE experiment design for LPV system identification and it is beyond the} %
scope {of} this paper.
  {In the reminder of the paper,} we will refer to $\dataset$ satisfying~\eqref{eq:thm:dim} as ``$\dataset$ being PE of degree~$(L,\dnx)$'' and we will consider that $\dataset$ satisfies this property throughout the paper, allowing us to} use representation~\eqref{eq:thm_FL} to analyze dissipativity of the considered LPV system in a fully data-driven setting.

\section{Data-driven dissipativity}
\label{sec:datadrivendissipativity}
In this section, we provide an equivalent characterization of dissipativity analysis of LPV systems based only on measured data.
First, in Section~\ref{subsec:datadrivendissipativity_finite}, we refine the notion of  $L$-dissipativity for LPV systems.
Section~\ref{subsec:datadrivendissipativity_LPV} then provides a necessary and sufficient condition for $L$-dissipativity of LPV systems based on data.

\subsection{Finite-horizon dissipativity}\label{subsec:datadrivendissipativity_finite}
As we consider finite amount of data, we will make use of %
Definition~\ref{def:L_diss}, requiring to satisfy the dissipation inequality over all time intervals up to a finite-time horizon $L$. The following result, based on \cite[Proposition 1]{romer2019one}, shows that it suffices to verify the dissipation inequality for only $L$.
\begin{prop}[$L$-dissipativity of LPV systems]\label{prop:L_diss}
The system defined by \eqref{eq:sys} is $L$-dissipative (see Definition \ref{def:L_diss}) w.r.t.~the supply rate~\eqref{eq:supplyrate} if
\begin{align}\label{eq:def_L_diss2}
\sum_{k=1}^{L}\begin{bmatrix}u_k\\y_k\end{bmatrix}^{\!\top} \!\!
\Pi\begin{bmatrix}u_k\\y_k\end{bmatrix}\geq0
\end{align}
for all trajectories $\{u_k,y_k\}_{k=1}^{L}\in\Bfint{\mr{IO}}{[1,L]}$ with $w_0=0$ (zero initial condition). %
\end{prop}
\begin{proof} 
As for any given ${p}_{[1,L]}\in\ms{P}_{[1,L]}$, the input-output behavior for LPV systems is \emph{linear}, therefore -- with the assumption of zero initial conditions -- the proof in \cite[Prop.~1]{romer2019one} can be {adopted} to prove this result. 
\end{proof}
Thus, to determine whether the LPV system~\eqref{eq:sys} is $L$-dissipative, we need to verify \eqref{eq:def_L_diss2} for all of its trajectories $(u_{[1,L]},y_{[1,L]})\in\Bfint{\mr{IO}}{[1,L]}$.
As we show next, the data-driven representation of \eqref{eq:sys}, which exists based on the Fundamental Lemma for LPV systems (see Proposition~\ref{prop:FLeasy}), allows us to translate this condition into a test. This test only depends on input-{scheduling}-output data and the set $\mathbb{P}$.

\begin{remark}[Finite- vs. infinite-horizon dissipativity]\label{rem:fin_vs_inf}
    In general, $L$-dissipativity is a more `optimistic' property than the infinite-horizon dissipativity property, given in Definition~\ref{def:dissipativity-hillmoylan}. %
{For} the case of SISO LTI systems, it is shown in %
\cite{boettcher2000toeplitz, tu2018approximation} that $L$-dissipativity with $\Pi$ defined with $(Q,S,R)=(\gamma^2, 0, -1)$ \emph{implies} (infinite-horizon) dissipativity with $\Pi$ defined as $(Q,S,R)=((\gamma+\epsilon)^2, 0, -1)$, where $\epsilon$ decays with~$\tfrac{1}{L^2}$, {i.e., the finite-horizon property converges to the infinite-horizon property}.
A similar statement holds true for more general dissipativity properties under suitable assumptions on a transformed LTI system and even extends to \emph{integral quadratic constraints}
(IQC) based analysis as shown in~\cite[Thm.~20]{koch2021determining}.
We conjecture that comparable bounds also apply to the relation between finite- and infinite-horizon dissipativity in the LPV case, which would result in data-driven dissipativity analysis that guarantees bounds on the performance of an unknown LPV system. {In} Section~\ref{ss:example2}, %
this decay of $\epsilon$, and thus convergence towards %
infinite-horizon dissipativity, %
{is illustrated} by means of an example.
\end{remark}
\begin{remark}
{In the LTI case, the infinite-horizon dissipativity property %
can also be analyzed via Willems' notion of dissipativity~\cite{Willems1972} by means of an extended state realization of the behavior~\cite{koch2021provably}, where the state is composed from lagged versions of $u$ and $y$. {However, this is difficult to obtain analogously in the LPV case, because the extended state realization for the behavior $\mf{B}$ represented by \eqref{eq:sys} results in an LPV-SS representation with \emph{dynamic} dependency. This invokes a complicated data-driven representation and a conservative dissipativity test, as it is based on an extended scheduling signal ($p_k$ and its lagged versions).}}
\end{remark}
\begin{remark}[Model-based $L$-dissipativity]
    It is rather trivial to work out a method to verify $L$-dissipativity in a model-based setting. For completeness, we give the derivation for $L$-dissipativity verification for LPV systems in Appendix~\ref{app:modelbasedLdissip}.
\end{remark}
{
\begin{remark}[Dissipativity analysis of generalized plants] 
    In model-based analysis, weighting filters are often employed to describe the expected performance specifications for a given control configuration by means of formulating a so-called \emph{generalized plant} form of the analysis problem on which the dissipativity check is performed. However, this requires interconnections of model-based elements and data-driven representations for which only recently a ``hybrid'' dissipativity approach has been developed in~\cite{Spin2023_DDGenPlant}. 
        The results of~\cite{Spin2023_DDGenPlant} %
        can also be applied {for the results of} this paper to derive data-driven dissipativity conditions {for} LPV generalized plants.
\end{remark}
}

\subsection{Data-driven dissipativity of LPV systems}\label{subsec:datadrivendissipativity_LPV}

Next, we derive an equivalent characterization of $L$-dissipativity for the unknown LPV system~\eqref{eq:sys} based on input-scheduling-output data. Given the \emph{data-dictionary} $\dataset = \{u_k, p_k, y_k\}_{k=1}^N\in\Bfint{}{[1,N]}$ measured from the system~\eqref{eq:sys}, we analyze $L$-dissipativity for length $L$ trajectories $(\bar{u}_{[1,L]}, \bar{p}_{[1,L]}, \bar{y}_{[1,L]})\in\Bfint{}{[1,L]}$. %

As in \cite{romer2019one}, we define the matrix $V_\tau\in\mathbb{R}^{(\dnu+\dny)\tau\times (\dnu+\dny)L}$ for some $\tau\in\mathbb{N}$ as 
\begin{equation}
	V_\tau = \begin{bmatrix}
 		I_{\tau\dnu} & 0_{\tau\dnu \times \dnu(L-\tau)} & 0_{\tau\dnu \times \dny \tau} & 0_{\tau\dnu \times \dny(L-\tau)} \\
		0_{\tau\dny  \times \tau\dnu} & 0_{\tau\dny \times \dnu(L-\tau)} & I_{\tau\dny} & 0_{\tau\dny \times \dny(L-\tau)}
 	\end{bmatrix}
\end{equation}
such that $V_\tau\begin{bmatrix} \bar{\vect{u}}_{L} \\ \bar{\vect{y}}_{L} \end{bmatrix}=0$ if and only if $\bar{u}_1=\dots=\bar{u}_{\tau}=0$ and $\bar{y}_1=\dots=\bar{y}_{\tau}=0$. {Having the condition $V_\tau\begin{bmatrix} \bar{\vect{u}}_{L} \\ \bar{\vect{y}}_{L} \end{bmatrix}=0$ next to the data-driven representation \eqref{eq:thm_FL} allows us to take only the trajectories of $\Bfint{}{[1,L]}$ with zero initial conditions into account %
{when} $\tau\geq\dnr$, which is required by Definition~\ref{def:L_diss}.}
 {Next,} let us define
\begin{subequations}\label{eq:Pi_H_F_p_def} 
\begin{align}
	\Pi_L  &= \begin{bmatrix}
		I_L\otimes Q & I_L\otimes S \\
		I_L\otimes S^\top & I_L\otimes R
	\end{bmatrix}, %
	\\
	\Pi_H  &= \begin{bmatrix}
		\mc{H}_L(u_{[1,N]}) \\ \mc{H}_L(y_{[1,N]})
	\end{bmatrix}^{\!\top}\!\! \Pi_L \begin{bmatrix}
		\mc{H}_L(u_{[1,N]}) \\ \mc{H}_L(y_{[1,N]})
	\end{bmatrix}, %
	\\
	F(\bar{p}_{[1,L]})  &= \begin{bmatrix}
		F_1 \\ F_2(\bar{p}_{[1,L]}) 
	\end{bmatrix} \nonumber\\
	&= \begin{bmatrix} 
		V_\tau \begin{bmatrix} \mc{H}_L(u_{[1,N]}) \\ \mc{H}_L(y_{[1,N]}) \end{bmatrix} \\
		\mc{H}_L(u_{[1,N]}^{\mt{p}})-\bar{\mathcal{P}}^{\dnu}\mc{H}_L(u_{[1,N]}) \vphantom{\Big)}\\
		\mc{H}_L(y_{[1,N]}^{\mt{p}})-\bar{\mathcal{P}}^{\dny}\mc{H}_L(y_{[1,N]}) \vphantom{\Big)}
	\end{bmatrix},
\end{align}
\end{subequations}
where $F(\bar{p}_{[1,L]})\in\mathbb{R}^{(\dnu+\dny)(\tau+\dnp L)\times (N-L+1)}$ {collects the restrictions on the IO behavior imposed by the scheduling signal and the requirement of having zero initial conditions}. Note that $F_2(\bar{p}_{[1,L]})$ contains both the scheduling trajectory ${p}_{[1,N]}$ of the \emph{data-dictionary} $\dataset$, {and} the scheduling trajectory\footnote{Note that $\bar{p}_{[1,L]}$ enters via $\bar{\mc{P}}^\bullet$.} $\bar{p}_{[1,L]}$ that is part of any length $L$ system trajectory in $\Bfint{}{[1,L]}$ for which we would like to test dissipativity. Fusing this with the results in~\cite{romer2019one} and~\cite{VerhoekTothHaesaertKoch2021}
leads us to the following result.
\begin{theorem}[$(L-\tau)$-dissipativity of LPV systems]\label{thm:LPV_diss}
The following statements hold:
\begin{enumerate}[label=(\roman*)]
\item \label{thm:LPV_diss:1} If the system defined by~\eqref{eq:sys} is $L$-dissipative {w.r.t.~a given $(Q,S,R)$}, then, for any {$\tau$ with} $\dnr\leq\tau<L$, and any $\bar{p}_{[1,L]}\in\ms{P}_{[1,L]}$, it holds that
\begin{align}\label{eq:thm_LPV_diss}
g^\top \Pi_H g \geq 0\quad\forall g\in\mathbb{R}^{N-L+1}:\>F(\bar{p}_{[1,L]})g=0.
\end{align}
\item \label{thm:LPV_diss:2} Conversely, suppose the system represented by~\eqref{eq:sys}  generates a PE data set $\dataset$ of order $(L,\dnx)$.
If there exists $\tau<L$ such that~\eqref{eq:thm_LPV_diss} holds for all $\bar{p}_{[1,L]}\in\ms{P}_{[1,L]}$, then {the system} is $(L-\tau)$-dissipative {w.r.t.~$(Q,S,R)$}.
\end{enumerate}
\end{theorem}
\begin{proof}
We first prove item~\ref{thm:LPV_diss:1}. 
Take an arbitrary $g$ satisfying $F(\bar{p}_{[1,L]})g=0$.
Then, Proposition~\ref{prop:FLeasy} implies 
that 
\begin{equation}\label{eq:thm1:traj}
	\begin{bmatrix} \bar{\vect{u}}_L \\ \bar{\vect{y}}_L \end{bmatrix} = \begin{bmatrix} \mc{H}_L(u_{[1,N]}) \\ \mc{H}_L(y_{[1,N]})\end{bmatrix} g 
\end{equation}
is a trajectory of~\eqref{eq:sys}. Moreover, due to $F_1g=0$, this trajectory satisfies $\bar{u}_1=\dots=\bar{u}_{\tau}=0$, $\bar{y}_1=\dots=\bar{y}_{\tau}=0$ 
 and since $\tau\geq \dnr$, {therefore} it has zero initial conditions.
Furthermore, since~\eqref{eq:sys} is $L$-dissipative {for $(Q,S,R)$}, the trajectory also necessarily satisfies
\begin{align}\label{eq:L_diss_stacked}
0 \leq \begin{bmatrix} \bar{\vect{u}}_L\\ \bar{\vect{y}}_L\end{bmatrix}^{\!\top} \!\! \Pi_L \begin{bmatrix} \bar{\vect{u}}_L\\ \bar{\vect{y}}_L \end{bmatrix}
=g^\top \Pi_H g,
\end{align}
i.e.,~\eqref{eq:thm_LPV_diss} holds.

We now prove item~\ref{thm:LPV_diss:2}. 
Let $(\bar{u}_{[\tau+1,L]},\bar{p}_{[\tau+1,L]},\bar{y}_{[\tau+1,L]})$ be an arbitrary trajectory in $\Bfint{}{[\tau+1,L]}$ with zero initial conditions.
We extend this trajectory by $\bar{u}_1=\dots=\bar{u}_{\tau}=0$ and $\bar{y}_1=\dots=\bar{y}_{\tau}=0$ such that $(\bar{u}_{[1,L]},\bar{p}_{[1,L]},\bar{y}_{[1,L]})$ for some $\bar{p}_{[1,\tau]}$ is also a trajectory in $\Bfint{}{[\tau+1,L]}$. %
Using Proposition~\ref{prop:FLeasy}, there exists a vector $g$ such that both \eqref{eq:thm1:traj} and %
$F(\bar{p}_{[1,L]})=0$ hold.
According to~\eqref{eq:thm_LPV_diss}, we then have $g^\top\Pi_H g\geq0$, i.e.,
\begin{align*}
0 \leq g^\top\Pi_H g=\begin{bmatrix} \bar{\vect{u}}_L\\ \bar{\vect{y}}_L \end{bmatrix}^{\!\top}\!\!\Pi_L \begin{bmatrix} \bar{\vect{u}}_L\\ \bar{\vect{y}}_L \end{bmatrix}
=\sum_{k=\tau+1}^{L}\begin{bmatrix}\bar{u}_k\\\bar{y}_k\end{bmatrix}^{\!\top}\!\!
\Pi
\begin{bmatrix}\bar{u}_k\\\bar{y}_k\end{bmatrix}.
\end{align*}
Since $\{\bar{u}_k,\bar{y}_k\}_{k=\tau+1}^{L}$ was arbitrary, this implies that~\eqref{eq:sys} is $(L-\tau)$-dissipative {for the considered $(Q,S,R)$}.
\end{proof}

Theorem~\ref{thm:LPV_diss} shows that~\eqref{eq:thm_LPV_diss} is a necessary and sufficient condition for $L$-dissipativity of the LPV system \eqref{eq:sys}.
Intuitively, the dissipation inequality $g^\top \Pi_Hg\geq0$ needs to hold for all vectors $g$ corresponding to zero initial conditions and the affine constraints in the third and fourth block row of~\eqref{eq:thm_FL}, i.e., satisfying $F(\bar{p}_{[1,L]})g=0$.
The proof follows similar arguments as the corresponding result for LTI systems~\cite[Thm.~2]{romer2019one}. {Furthermore, note that this result is in-fact recovered when $F_2$ in~\eqref{eq:Pi_H_F_p_def} is set to zero, which is the case for a constant scheduling, i.e., $p_k=\bar{p}_k=c\in\mb{P}$ for all $k$, resulting in an LTI system.}

{As $N$ is determined by the PE condition, %
the hyperparameters for the $(L-\tau)$-dissipativity analysis are $L$ and $\tau$. {If $\tau<\dnr$, then (ii) in Theorem~\ref{thm:LPV_diss} is typically not satisfied~\cite{romer2019one}. Hence, $\tau$ is often %
 chosen large enough to robustly satisfy this condition, e.g., twice of the expected system lag.  
 For choosing $L>\tau$, the conjectured convergence for $\ell_2$-gain in Remark~\ref{rem:fin_vs_inf} implies that taking it as large as possible, brings the results closer to the infinite-horizon dissipativity of the original system. However, this requires large enough $N$ to support this choice and, as we will see, it also increases computational complexity of testing  $(L-\tau)$-dissipativity.} 
   We further comment on choosing the hyperparameters $L$ in Section~\ref{sec:examples}.}

\section{Computational approaches for verifying data-driven dissipativity}
\label{sec:computational}
In this section,  we derive tractable conditions for the verification of dissipativity properties of LPV systems based on data. To this end, we can observe that verifying~\eqref{eq:thm_LPV_diss} can be seen as a robust optimization problem~\cite{ben2009robust}. Hence, we will solve the data-driven analysis problem of Theorem~\ref{thm:LPV_diss} by applying Finsler's Lemma, which translates it into a feasibility check of a scheduling-dependent definiteness condition for all $\bar p_{[1,L]}\in\ms{P}_{[1,L]}$. While this feasibility check involves verifying infinitely many definiteness conditions (one for every trajectory in $\ms{P}_{[1,L]}$), we will show that it can be reduced to a finite-dimensional LMI feasibility problem by taking structural assumptions on $\ms{P}_{[1,L]}$ and extending the analysis results from the model-based setting~\cite{apkarian2000parameterized, HoffmannWerner2014, parrilo2000structured, oliveira2007parameter}.

\subsection{Main concept}

{To derive a computable form of the data-driven analysis problem of Theorem~\ref{thm:LPV_diss},} %
we introduce the \emph{non-strict version} of Finsler's lemma~\cite{meijer2024general}
{that will prove to be an essential building block of our proposed approaches.}
\begin{lemma}[{Non-strict Finsler's Lemma \cite{meijer2024general}}]\label{lem:Finsler}
Let $A\in\mathbb{R}^{n\times n}$, $B\in\mathbb{R}^{q\times n}$.
Then, the following statements: %
\begin{enumerate}[label={(\roman*)}]
\item $x^\top Ax\geq0$ for all $x$ satisfying $Bx=0$,

\item $(B^\perp)^\top AB^\perp\succeq0$,

\item There exists a $\mu\in\mathbb{R}$ such that $A+\mu B^\top B\succeq0$,

\item There is an $X\in\mathbb{R}^{n\times q}$ such that $XB+B^\top X^\top - A \preceq 0$,
\end{enumerate}
satisfy that ``(i) $\Leftrightarrow$ (ii)'', ``(iii) $\Rightarrow$ (ii)'', and ``(iv) $\Rightarrow$ (ii)''.
\end{lemma}
According to Theorem~\ref{thm:LPV_diss}, analyzing dissipativity amounts to verifying Condition~\eqref{eq:thm_LPV_diss} for all {admissible} scheduling signals $\bar{p}_{[1,L]}\in\ms{P}_{[1,L]}$. Based on ``(i) $\Leftrightarrow$ (ii)'' in Lemma~\ref{lem:Finsler}, we infer that~\eqref{eq:thm_LPV_diss} holds if and only if
\begin{align}\label{eq:dd_Ldiss_Finsler}
\big(F(\bar{p}_{[1,L]})^\perp\big)^\top \Pi_H F(\bar{p}_{[1,L]})^\perp\succeq0.
\end{align}
The same idea has been been employed in the LTI case in~\cite{romer2019one}, where this reformulation 
provides a simple and elegant condition to verify dissipativity, only requiring to check positive semi-definiteness of a given data-dependent matrix.
For LPV systems of the form \eqref{eq:sys}, condition~\eqref{eq:dd_Ldiss_Finsler} allows us to check whether the dissipation inequality holds for a \emph{fixed} scheduling trajectory $\bar{p}_{[1,L]}$. However, we need to verify dissipativity as a \emph{system} property for the whole $\ms{P}_{[1,L]}$, which makes, together with the fact that
$F(\bar{p}_{[1,L]})^\perp$ depends nonlinearly on $\bar{p}_{[1,L]}$, the verification of~\eqref{eq:dd_Ldiss_Finsler} intractable.

Therefore, in the remainder of this section, we exploit items~(iii) and~(iv) of Lemma~\ref{lem:Finsler} in order to derive computational procedures for the verification of~\eqref{eq:thm_LPV_diss}, which are tailored to the assumed {definition of $\ms{P}_{[1,L]}$}.
First, in Section~\ref{subsec:computational_polytopic}, we employ item (iv) to address {the case when} $\mathbb{P}$ {is a convex, polytopic set, by describing} $\ms{P}_{[1,L]}$ based on a convex hull argument.
Next, in Section~\ref{subsec:computational_S_procedure}, we use the S-procedure to handle admissible scheduling sets $\ms{P}_{[1,L]}$ defined by quadratic inequalities.

\subsection{Dissipativity verification via a convex hull argument}\label{subsec:computational_polytopic}
\subsubsection{Core concept} We now provide a computational procedure to verify~\eqref{eq:thm_LPV_diss} for any $\bar{p}_{[1,L]}\in\ms{P}_{[1,L]}$, where $\ms{P}_{[1,L]}:=\mb{P}^L$ and $\mathbb{P}$ is a %
convex polytope {with finite many vertices}.

\begin{assumption}[{Polytopic description of $\mb{P}$}]\label{ass:polytopic}
The set $\mathbb{P}\subset\mb{R}^{\dnp}$ is a convex polytope, {generated by a finite {number} of vertices, i.e.,   %
$\mathbb{P}=\mathrm{co}(\{\msf{p}_i\}_{i=1}^{n_\mathrm{v}})$, where $\mathrm{co}$ denotes the convex hull.} \hfill$\square$
\end{assumption}
As the scheduling variable varies in $\mb{P}$, i.e., {$\forall k \in\mathbb{Z}:\bar{p}_k\in\mb{P}$}, every trajectory {$\bar{p}_{[1,L]}$} is confined in the space 
\[\ms{P}_{[1,L]}=\mb{P}\times\dots\times\mb{P}=\mb{P}^L,\]
which is generated with $n_\mathrm{v}^L$ vertices {that} are all permutations of the original vertices of $\mb{P}$, i.e., $\ms{P}_{[1,L]}= \mathrm{co}( \mathrm{perm}_L(\{\msf{p}_i\}_{i=1}^{n_\mathrm{v}}) )$.
We now apply ``(iv) $\Rightarrow$ (i)'' in Lemma~\ref{lem:Finsler} to derive a tailored condition for verifying~\eqref{eq:thm_LPV_diss} via a convex hull argument, i.e., by means of Assumption~\ref{ass:polytopic}. {For this, we will explicitly write $F(\bar{p}_{[1,L]})$ in its affine form, i.e., with $\bar{\vect{p}}_{[1,L]}\in\mb{R}^{L\dnp}$,
\begin{equation}\label{eq:affineF}
    F(\bar{p}_{[1,L]}) = \hat{F}_0 + \sum_{i=1}^{L\dnp} \bar{\vect{p}}_{i}\hat{F}_i. %
\end{equation}
}

\begin{prop}[$L$-dissipativity via convex hull argument]\label{prop:LPV_diss_polytopic}
Suppose Assumption~\ref{ass:polytopic} holds and let $\{\bar{\msf{p}}_\nu\}_{\nu=1}^{n_\mathrm{v}^L}$ denote the vertices of $\mb{P}^L$. 
Then,~\eqref{eq:thm_LPV_diss} holds {if} %
there exist matrices $X_i\in\mathbb{R}^{(N-L+1)\times(N-L+1)}$, {$i \in \{ 0,\dots,L\dnp\}$},
\begin{subequations}\label{eq:prop_LPV_diss_polytopic}
satisfying{
\begin{align}
    \sum_{i=0}^{L\dnp}\sum_{j=0}^{L\dnp} \bar{\vect{p}}_{i}\bar{\vect{p}}_{j}\left(X_i \hat{F}_j + \hat{F}_j^\top X_i^\top\right) -\Pi_H\preceq0, \label{eq:prop_LPV_diss_polytopic:a}\\
    X_i \hat{F}_i + \hat{F}_i^\top X_i^\top \succeq 0, \quad {\forall i \in \{ 0,\dots,L\dnp \}},\label{eq:prop_LPV_diss_polytopic:b}
\end{align}}%
\end{subequations}
{for all vertices {$\{ \bar{\msf{p}}_\nu\}_{\nu=1}^{n_\mathrm{v}^L}$}.}
\end{prop}
\begin{proof}
For any given scheduling trajectory $\bar{p}_{[1,L]}\in\ms{P}_{[1,L]}$, Lemma~\ref{lem:Finsler} implies that~\eqref{eq:thm_LPV_diss} holds if there exists {a} matrix function $X(\bar{p}_{[1,L]}):\ms{P}_{[1,L]} \to \mathbb{R}^{(N-L+1)\times(N-L+1)}$ associated with $\bar{p}_{[1,L]}$ such that
\begin{align}\label{eq:dd_Ldiss_Finsler1}
	X(\bar{p}_{[1,L]})F(\bar{p}_{[1,L]})+F(\bar{p}_{[1,L]})^\top X(\bar{p}_{[1,L]})^\top\!\! -\Pi_H \preceq 0.
\end{align}
{By considering $X$ to have affine dependency on $\bar{p}_{[1,L]}$, i.e., $X(\bar{p}_{[1,L]})=X_0+\sum_{i=1}^{L\dnp} \bar{\vect{p}}_{i}X_i$,  \eqref{eq:dd_Ldiss_Finsler1} under~\eqref{eq:affineF} reads as
\begin{equation}\label{eq:pf:nonconvex}
    \sum_{i=0}^{L\dnp}\sum_{j=0}^{L\dnp} \bar{\vect{p}}_{i}\bar{\vect{p}}_{j}\left(X_i \hat{F}_j + \hat{F}_j^\top X_i^\top\right) -\Pi_H\preceq0, \ \forall \bar{p}_{[1,L]}\in\ms{P}_{[1,L]},
\end{equation}
which {is quadratic, but not necessarily convex in $\bar{p}_{[1,L]}$}}. %
Based on the application of the multi-convexity relaxation from~\cite{apkarian2000parameterized, gahinet1996affine}, condition~\eqref{eq:prop_LPV_diss_polytopic:b} enforces convexity of~\eqref{eq:pf:nonconvex}, and hence if~\eqref{eq:prop_LPV_diss_polytopic:a} and~\eqref{eq:prop_LPV_diss_polytopic:b} hold for all vertices $\{ \bar{\msf{p}}_\nu\}_{\nu=1}^{n_\mathrm{v}^L}$ of $\mathbb{P}^L$, then~\eqref{eq:pf:nonconvex} holds for all $\bar{p}_{[1,L]}\in\ms{P}_{[1,L]}$.
\end{proof}

Proposition~\ref{prop:LPV_diss_polytopic} provides a computational approach to verify condition~\eqref{eq:thm_LPV_diss} in case the scheduling variable is varying in a convex polytope (Assumption~\ref{ass:polytopic}).
Thus, we have reduced the {condition of} data-driven dissipativity %
in Theorem~\ref{thm:LPV_diss}, {which needs to be checked for all $\bar{p}_{[1,L]}\in\ms{P}_{[1,L]}$,}  %
to the existence of matrices $X_i$ such that~\eqref{eq:prop_LPV_diss_polytopic} holds, i.e., to an SDP subject to LMI constraints.  %
 
\subsubsection{{Including rate bounds on $\mathit{p}$}}\label{sss:conservatism}
{We now discuss the inclusion of additional system properties into the analysis problem in terms of incorporating rate bounds on the scheduling signal.}
The analysis in Proposition~\ref{prop:LPV_diss_polytopic} allows for maximal \emph{variation} of the scheduling variable, i.e., $p_k-p_{k-1}$ is only limited to remain inside $\mb{P}$. This can result in conservative conclusions, e.g., on the $\ell_2$-gain of the LPV system, as the analysis also considers (possibly non-existent) fast variations of the scheduling signal. We can reduce this %
by {incorporating} %
rate bounds on $p$, {i.e.,} %
defining the {admissible scheduling set as %
\begin{equation}
    {\ms{P}_{[1,L]}} %
    = \{ p_{[1,L]}\in\mb{P}^L \mid p_k-p_{k-1}\in\mb{D}, \ \forall k = 2, \dots, L \},
\end{equation}
where $\mb{P}$ satisfies Assumption~\ref{ass:polytopic}} and $\mb{D}$ is a convex polytope that defines the rate bounds on $p$. {Note that {$\ms{P}_{[1,L]}$}  is again a convex polytope.} By verifying \eqref{eq:prop_LPV_diss_polytopic} on the vertices of {$\ms{P}_{[1,L]}$}, %
{the analysis yields a %
less conservative
conclusion on the dissipativity property of the system.}
\begin{remark}[Computational complexity vs. conservatism]\label{rem:manylmis}
	Verifying dissipativity via \eqref{eq:prop_LPV_diss_polytopic} \emph{without} incorporation of the rate bounds on $p$ requires solving an SDP with {$(1+L\dnp)n_\mathrm{v}^L$} LMI constraints. This number further increases when including rate bounds, which increases the computational complexity exponentially for larger $L$. Explosion of the computational load for large $L$ can be mitigated by replacing~\eqref{eq:prop_LPV_diss_polytopic} with
\[ XF(\bar{\msf{p}}_\nu) + F(\bar{\msf{p}}_\nu)^\top X^\top  -\Pi_H \preceq0, \quad \nu=1,\dots, n_\mr{v}^L.\]
Introducing a constant $X$ can introduce conservatism, however, it also allows to alleviate the computational load of the dissipativity test.
\end{remark}
{We will see in Section~\ref{sec:examples} that Proposition~\ref{prop:LPV_diss_polytopic} is readily applicable to small-scale LPV systems and it provides tight conditions for verifying $L$-dissipativity from data.}

\subsection{Dissipativity verification via the S-procedure}\label{subsec:computational_S_procedure}
In this section, we assume that $\bar{p}_{[1,L]}$ is varying in a space that is described by a quadratic inequality instead of a convex polytope, i.e., we assume that the scheduling signals in $\ms{P}_{[1,L]}$ are described as follows:
\begin{assumption}[{Quadratic description of $\ms{P}_{[1,L]}$}]\label{ass:ellipsoidal}
The scheduling signal $\bar{p}_{[1,L]}$ varies around a nominal scheduling trajectory $\breve{p}_{[1,L]}$, such that $\bar{p}_{[1,L]}$ satisfies
\begin{align}\label{eq:p_norm_bound}
\lVert \bar{p}_k-\breve{p}_{k}\rVert_W\leq p_{\max}
\end{align}
for all $k=1,\dots,L$ with some $p_{\max}>0$. The set $\ms{P}_{[1,L]}$ can then be trivially described by  
\begin{equation}\label{eq:ass_ellipsoidal}
	{\ms{P}_{[1,L]}}=\left\{\bar{p}_{[1,L]}\in\mb{P}^L \Bigm| \vphantom{ \begin{bmatrix} \hat{\mc{P}}^{\dnu+\dny} \\ I \end{bmatrix}^\top} %
	\begin{bmatrix} \bar{\mc{P}}^{\dnu,\dny} \\ I \end{bmatrix}^{\!\top}\!\! M_P \begin{bmatrix} \bar{\mc{P}}^{\dnu,\dny} \\ I \end{bmatrix}\succeq 0 \right\}, %
\end{equation}
with $\bar{\mc{P}}^{\dnu,\dny}=\begin{bmatrix} \bar{\mathcal{P}}^{\dnu} & 0 \\ 0 & \bar{\mathcal{P}}^{\dny}\end{bmatrix}$, $\bar{\mathcal{P}}^{n_\bullet}=\bar{p}_{[1,L]}\bkron I_{n_\bullet}$.
\hfill$\square$
\end{assumption}
{Note that for $W=2$ in~\eqref{eq:p_norm_bound},} %
we have that $M_P$ in~\eqref{eq:ass_ellipsoidal} is
\begin{align}\label{eq:def_of_MP_ellipsiodal}
	M_P = \begin{bmatrix}
		-I & \breve{\mathcal{P}}^{\dnu,\dny} \\
		(\breve{\mathcal{P}}^{\dnu,\dny})^\top & p_{\max}^2 I -(\breve{\mathcal{P}}^{\dnu,\dny})^\top\breve{\mathcal{P}}^{\dnu,\dny}
	\end{bmatrix},
\end{align}
with $\breve{\mc{P}}^{\dnu,\dny} =\begin{bmatrix}\breve{\mc{P}}^{\dnu}&0\\0&\breve{\mc{P}}^{\dny}\end{bmatrix}$, $\breve{\mc{P}}^{n_\bullet}=\breve{p}_{[1,L]}\bkron I_{n_\bullet}$. Note that with Assumption~\ref{ass:ellipsoidal}, we can describe types of $\ms{P}_{[1,L]}$ that have a spherical or {hyper-rectangular} form. %
Next, we employ the S-procedure (see, e.g.,~\cite{scherer1997full, scherer2001lpv}) to derive a tractable reformulation of~\eqref{eq:thm_LPV_diss} for scheduling sets $\ms{P}_{[1,L]}$ that satisfy Assumption~\ref{ass:ellipsoidal}.
To this end, let us write $F(\bar{p}_{[1,L]})$ in~\eqref{eq:Pi_H_F_p_def} as
\begin{equation*}
\hspace{-0.5mm}	F(\bar{p}_{[1,L]})\!  =\!\!  %
	\underbrace{\begin{bmatrix} F_1 \\ \mc{H}_L(u_{[1,N]}^{\mt{p}}) \\ \mc{H}_L(y_{[1,N]}^{\mt{p}}) \end{bmatrix}}_{F_3}\!  -\! 	\underbrace{\begin{bmatrix} 0 & 0 \\ I & 0 \\ 0 & I \end{bmatrix}}_{F_4}\!\! \begin{bmatrix} \bar{\mathcal{P}}^{\dnu}\!\!\! & 0 \\ 0 & \!\!\!\bar{\mathcal{P}}^{\dny}\end{bmatrix} \!\! \underbrace{\begin{bmatrix} \mc{H}_L(u_{[1,N]}) \\ \mc{H}_L(y_{[1,N]}) \end{bmatrix}}_{F_5}\!.
\end{equation*}
\begin{prop}[$L$-dissipativity via the S-procedure]\label{prop:LPV_diss_ellipsoidal}
Suppose Assumption~\ref{ass:ellipsoidal} holds.
Then,~\eqref{eq:thm_LPV_diss} holds if there exist a $\mu\in\mathbb{R}$ and a $\lambda\geq0$ such that
\begin{align}\label{eq:prop_LPV_diss_ellipsoidal}
\begin{bmatrix}\mu F_4^\top F_4&\mu F_4^\top F_3\\
\mu F_3^\top F_4&\mu F_3^\top F_3+\Pi_H\end{bmatrix}-\lambda \begin{bmatrix}I&0\\0&F_5\end{bmatrix}^{\!\top} \!\!
M_P
\begin{bmatrix}I&0\\0&F_5\end{bmatrix}\succeq0.
\end{align}
\end{prop}

\begin{proof}
{Using}
the implication of (i) by (iii) in Lemma~\ref{lem:Finsler},~\eqref{eq:thm_LPV_diss} holds if for each $\bar{p}_{[1,L]}\in\ms{P}_{[1,L]}$ there exists a $\mu\in\mathbb{R}$ satisfying
\begin{align}\label{eq:prop_LPV_diss_ellipsoidal_proof1}
	\Pi_H+\mu F^\top\!(\bar{p}_{[1,L]}) F(\bar{p}_{[1,L]})\succeq0.
\end{align}
 {Note that {in terms of this condition,} $\mu$ {can be different for each} $\bar{p}_{[1,L]}$. To obtain a convex problem, we {enforce} $\mu$ to be {the same}\footnote{{See Remark~\ref{rem:PVmulti} for a discussion on a scheduling dependent $\mu$.}} for all $\bar{p}_{[1,L]}\in\ms{P}_{[1,L]}$}, {which inherently introduces conservatism.}
{Then, we can rewrite}~\eqref{eq:prop_LPV_diss_ellipsoidal_proof1} {as} %
\begin{align}\label{eq:prop_LPV_diss_ellipsoidal_proof2}
\begin{bmatrix}\bar{\mathcal{P}}^{\dnu,\dny}F_5\\I\end{bmatrix}^\top
\begin{bmatrix}\mu F_4^\top F_4&\mu F_4^\top F_3\\
\mu F_3^\top F_4&\mu F_3^\top F_3+\Pi_H\end{bmatrix}
\begin{bmatrix}\bar{\mathcal{P}}^{\dnu,\dny}F_5\\I\end{bmatrix}\succeq0.
\end{align}
{By} left and right {multiplication of {the}} inequality in~\eqref{eq:ass_ellipsoidal} {with} $F_5^\top$ and $F_5$, respectively, we infer that $\bar{p}_{[1,L]}\in\ms{P}_{[1,L]}$ {if and only if} %
\begin{align}\label{eq:prop_LPV_diss_ellipsoidal_proof3}
	\begin{bmatrix} \bar{\mathcal{P}}^{\dnu,\dny}F_5 \\ I \end{bmatrix}^\top
	\begin{bmatrix} I & 0 \\ 0 & F_5 \end{bmatrix}^\top
	M_P
	\begin{bmatrix} I & 0 \\ 0 & F_5 \end{bmatrix}
	\begin{bmatrix} \bar{\mathcal{P}}^{\dnu,\dny}F_5 \\ I \end{bmatrix}\succeq 0.
\end{align}
Multiplying~\eqref{eq:prop_LPV_diss_ellipsoidal} with $\begin{bmatrix}(\bar{\mathcal{P}}^{\dnu,\dny}F_5)^\top & I \end{bmatrix}$ from the left and with $\begin{bmatrix}(\bar{\mathcal{P}}^{\dnu,\dny}F_5)^\top & I \end{bmatrix}^\top$ from the right and using~\eqref{eq:prop_LPV_diss_ellipsoidal_proof3}, we {get that if \eqref{eq:prop_LPV_diss_ellipsoidal} holds, then}~\eqref{eq:prop_LPV_diss_ellipsoidal_proof2} is implied, which concludes the proof.
\end{proof}

Proposition~\ref{prop:LPV_diss_ellipsoidal} provides a sufficient condition for~\eqref{eq:thm_LPV_diss} and hence, by Theorem~\ref{thm:LPV_diss}, for $(L-\tau)$-dissipativity of the underlying LPV system.
Verifying feasibility of~\eqref{eq:prop_LPV_diss_ellipsoidal} is an SDP with only two decision variables $\mu$ and $\lambda$, and thus it can be {efficiently} implemented from a {computational} perspective for {moderate and even large choices of $L$ and $N$} %
data lengths (see Section~\ref{sec:examples} for a numerical example).

Note that, {similar to} Proposition~\ref{prop:LPV_diss_polytopic}, Proposition~\ref{prop:LPV_diss_ellipsoidal} only provides a \emph{sufficient} condition for~\eqref{eq:thm_LPV_diss}.
This is because the multiplier $\mu\in\mathbb{R}$ %
is chosen {to be the same for all} %
scheduling {trajectories} $\bar{p}_{[1,L]}$.
Therefore, if, e.g., a norm bound similar to~\eqref{eq:p_norm_bound} is available, then translating this bound into a convex polytope as in Assumption~\ref{ass:polytopic} and applying Proposition~\ref{prop:LPV_diss_polytopic} will %
generally lead to less conservative results than considering the quadratic description in Assumption~\ref{ass:ellipsoidal} and applying Proposition~\ref{prop:LPV_diss_ellipsoidal}, even for a constant $X$.
On the other hand, as we will also demonstrate with a numerical example in Section~\ref{sec:examples}, Proposition~\ref{prop:LPV_diss_ellipsoidal} is significantly more efficient  from a computational perspective than Proposition~\ref{prop:LPV_diss_polytopic}.

{Finally, we want to highlight that for all the introduced computational methods, we assume that the `true' admissible scheduling set $\ms{P}_{[1,L]}$ is equivalent with the assumed descriptions (Assumptions~\ref{ass:polytopic}~and~\ref{ass:ellipsoidal}). If the assumed descriptions are in fact \emph{over approximating} $\ms{P}_{[1,L]}$, which can happen when the LPV representation is in fact a \emph{surrogate} model for a nonlinear system, then this again introduces a source of conservatism in the analysis. This issue of conservatism and minimizing it is actively studied in %
constructing LPV surrogate models of nonlinear systems \cite{SADEGHZADEH20204737}.}

\begin{remark}[{Sampling-based $L$-dissipativity analysis}]\label{rem:PVmulti}
The described conservatism of Proposition~\ref{prop:LPV_diss_ellipsoidal} can be alleviated by considering \emph{parameter-dependent} multipliers for the application of Finsler's Lemma (Lemma~\ref{lem:Finsler}), similar to Proposition~\ref{prop:LPV_diss_polytopic}.
To be precise,~\eqref{eq:prop_LPV_diss_ellipsoidal_proof1} can be replaced by
\begin{align}\label{eq:prop_LPV_diss_ellipsoidal_pv}
	\Pi_H+\mu(\bar{p}_{[1,L]}) F(\bar{p}_{[1,L]})^\top F(\bar{p}_{[1,L]})\succeq0,
\end{align}
for some $\mu:\ms{P}_{[1,L]}\to\mathbb{R}$.
In fact, by applying Finsler's Lemma point-wise for any parameter trajectory $\bar{p}_{[1,L]}\in\ms{P}_{[1,L]}$, the existence of a mapping $\mu(\cdot)$ such that~\eqref{eq:prop_LPV_diss_ellipsoidal_pv} holds is a less conservative\footnote{In fact, the provided condition is also necessary, and therefore non-conservative, for \emph{strict} $L$-dissipativity.\label{footnote}} condition for~\eqref{eq:thm_LPV_diss} than~\eqref{eq:prop_LPV_diss_ellipsoidal}.
In practice, inequality~\eqref{eq:prop_LPV_diss_ellipsoidal_pv} can be checked by drawing $N_\mathrm{s}$ (e.g., uniformly distributed) samples $\hat{p}_{[1,L]}^i$ from $\ms{P}_{[1,L]}$ and verifying the existence of $\mu_i\in\mathbb{R}$, $i=1,\dots,N_\mathrm{s}$ such that
\begin{align}
	\Pi_H+\mu_i F(\hat{p}_{[1,L]}^i)^\top F(\hat{p}_{[1,L]}^i)\succeq0
\end{align}
for all $i=1,\dots,N_\mathrm{s}$.
For any fixed number $N_\mathrm{s}$, this yields a less conservative\tss{\ref{footnote}} condition for~\eqref{eq:thm_LPV_diss} and, hence, for $L$-dissipativity compared to Proposition~\ref{prop:LPV_diss_ellipsoidal}.
The resulting dissipativity test is \emph{tight} in the limit $N_\mathrm{s}\to\infty$.
While choosing a dense grid over $\ms{P}_{[1,L]}$, i.e., a large value of $N_\mathrm{s}$, can be prohibitive due to the curse of dimensionality, good results can often be achieved in practice for reasonable values of $N_\mathrm{s}$, see the numerical example in Section~\ref{sec:examples}.
\end{remark}
\begin{remark}[Reducing conservatism with matrix-valued multipliers]
    Another option to reduce the conservatism of Proposition~\ref{prop:LPV_diss_ellipsoidal} is to consider a matrix-valued~$\lambda$. As \eqref{eq:p_norm_bound} constitutes~$L$ inequalities, we can %
    {introduce} at least~$L$ degrees of freedom in~$\lambda$ using %
    full-block multipliers~\cite{scherer1997full}. {Using full-block multipliers} does, on the other hand, significantly increase the complexity of the problem. Exploring such a formulation of the analysis is interesting for further research.
\end{remark}

\section{Examples}
\label{sec:examples}
This section demonstrates the effectiveness of the developed theory by means of determining the $\ell_2 $-gain of the considered LPV systems {in two academic examples} {and one realistic case-study on the LPV embedding of a nonlinear unbalanced disc system}. All the computations discussed in this section have been executed using \matlab on a MacBook Pro~(2020) with Intel Core i5 processor and solved using YALMIP \cite{YALMIP} with the MOSEK solver \cite{MOSEK}.

\subsection{Example I: {Model-based vs. data-driven analysis}} \label{example:1}
\begin{figure}
\centering
\includegraphics[scale=1]{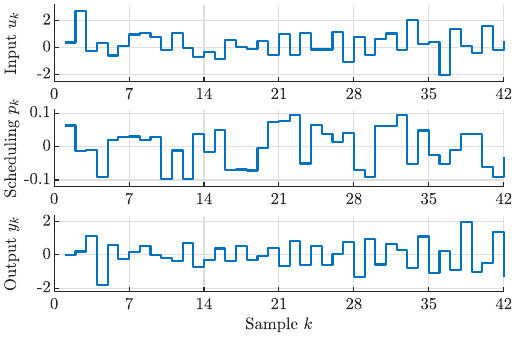} %
\caption{Data-dictionary for Example I.}\label{fig:datadictionary} %
\end{figure}

In this example, we compare our direct data-driven {analysis }approaches with a model-based and an indirect data-driven\footnote{With indirect data-driven method we mean to first perform system identification that results in a model on which model-based analysis {is performed}.} method. Consider an LPV system {described in the form of} \eqref{eq:sys} with $\dny=\dnu=\dnp=1$, $\dna=\dnb=3$ and $a,b$  as
\begin{align*}
    \begin{bmatrix} a_{1,0} & a_{2,0} & a_{3,0} \\ a_{1,1} & a_{2,1} & a_{3,1} \end{bmatrix} & =%
    \begin{bmatrix} 0.0826    & 0.1491    & -0.1196   \\ 0.0311    & 0.0570    & -0.0650   \end{bmatrix}, \\
    \begin{bmatrix} b_{1,0} & b_{2,0} & b_{3,0} \\ b_{1,1} & b_{2,1} & b_{3,1} \end{bmatrix} & =%
    \begin{bmatrix} 0.5007    & -0.5588   & 0.0784    \\ 0.6953    & 1.8192    & -1.7192   \end{bmatrix}.
\end{align*}
Moreover, we define $\mb{P}$ as $[-0.1,\> 0.1]$ with no rate bounds on~$p$, i.e., $\ms{P}_{[1,L]}=[-0.1,\> 0.1]^L$. For our data-driven dissipativity analysis, we want to determine the $\ell_2 $-gain of this system for a horizon of 7. Hence, we choose $\tau=\max\{\dna,\dnb\}=3$ and set $L$ to 10. {As the system is {SISO}, we have that $\dnx=\max\{\dna,\dnb\}={n_\mathrm{r}}= 3$ \cite{Toth2011_SSrealizationTCST}.} By exciting the system with a {white noise} input $u_k\sim\mc{N}(0,1)$ and scheduling $p_k\sim\mc{U}(\mb{P})$, {where $\mc{N}$ and $\mc{U}$ denote normal and uniform distributions, respectively}, the data-dictionary $\dataset$  with $N=42$ shown in Fig.~\ref{fig:datadictionary} is persistently exciting of order $(L,\dnx)$ {in terms of Proposition~\ref{prop:FLeasy}}.
{We compute an upper bound $\gamma$ on the $(L-\tau)$-horizon $\ell_2 $-gain with the model-based approach that is %
{given}
in Appendix~\ref{app:modelbasedLdissip}. We verify~\eqref{eq:LMImodelbased} by using a polytopic description of $\ms{P}_{[1,L]}$ and solving~\eqref{eq:LMImodelbased} on the vertices of $\ms{P}_{[1,L]}$, while minimizing the upper bound $\gamma$. Solving this problem yields  $\gamma=1.362$.} 
Next, we employ the \lpvcore toolbox\footnote{See \cite{BoefCoxToth2021} and \texttt{lpvcore.net} for {this} open-source \matlab toolbox.} to identify {a state-space model of} the LPV system using $\dataset$ and compute the $(L-\tau)$-horizon $\ell_2 $-gain of the resulting identified model. For the identification, we use {the} LPV PEM-SS {method} (see \cite{Toth18cAUT, Cox2018})
with the default settings in \lpvcore~{via} the function \texttt{lpvssest}. The {resulting}  {upper bound $\gamma$ on the $(L-\tau)$-horizon} $\ell_2 $-gain of the identified {LPV} model is 
1.375.%

Finally, we determine the $(L-\tau)$-horizon $\ell_2 $-gain of the system \emph{directly} with $\dataset$ using the results in this paper {to demonstrate} %
 that our direct data-driven dissipativity verification methods are competitive with their (indirect) model-based counter-parts. First, we employ the direct data-driven verification method via the convex hull argument. With the defined range for $p$, we construct the hypercube $\mb{P}^L$  {that defines $\ms{P}_{[1,L]}$}. The resulting hypercube has $n_\mathrm{v}=1024$ vertices. Together with the construction of $F$ using \emph{only} the data-dictionary $\dataset$, we can now solve~\eqref{eq:prop_LPV_diss_polytopic} with a constant $X$ on the vertices of $\mb{P}^L$ and minimize the value $\gamma$ that corresponds to an  {upper bound on the $(L-\tau)$-horizon} $\ell_2 $-gain of the system. The corresponding SDP yields $\gamma=1.362$,  {which demonstrates equivalence with the model-based method}. For the direct data-driven verification method using the S-procedure, we again use $\dataset$ to construct the necessary matrices to apply Proposition~\ref{prop:LPV_diss_ellipsoidal}. Note that by the definition of the scheduling region,  {we can equivalently describe $\ms{P}_{[1,L]}$ using \eqref{eq:def_of_MP_ellipsiodal} by choosing $p_\mr{max}=0.2$ and $\breve{p}_{[1,L]}$ as a zero trajectory.} Solving~\eqref{eq:prop_LPV_diss_ellipsoidal} yields a value for $\gamma$ of 1.831, representing a bound on the $(L-\tau)$-horizon $\ell_2 $-gain of the LPV system. This illustrates the aforementioned conservatism of Proposition~\ref{prop:LPV_diss_ellipsoidal}. On the other hand, the computational load of the latter method is significantly smaller, which we will further illustrate in Example~II, after demonstrating the influence of noise on the data-driven dissipativity analysis.

\subsubsection*{Influence of noisy data}
\begin{figure}
\centering
\includegraphics[scale=1]{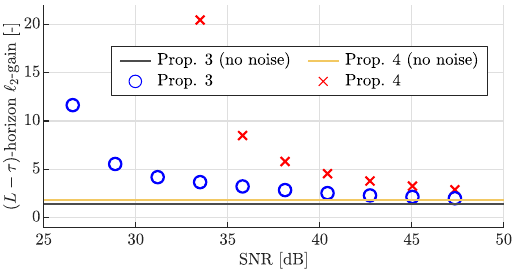} 
\caption{Influence of noisy data on the $(L-\tau)$-dissipativity analysis.}\label{fig:influencenoise} 
\end{figure}

To study the influence of noise on the $L$-dissipativity analysis, {we repeat} the analysis of this example for different levels of noise injected into the system {in terms of an ARX-type of structure:
\begin{align}\label{eq:noisy}
y_k+\sum_{i=1}^{\dna}a_i(p_{k-i})y_{k-i}=\sum_{i=0}^{\dnb}b_i(p_{k-i})u_{k-i} + e_k,
\end{align} 
where $e_k$ is an i.i.d. white noise process with $e_k\in\mathcal{N}(0,\sigma_\mathrm{e}^2)$.}
 For every analysis experiment, we use the same noise realization (sampled from a normal distribution) and adjust the variance of the noise by means of multiplication {of the sequence} with a constant. {In this way,} we obtain data-dictionaries with a \emph{signal-to-noise}~(SNR) between 26 and 47 dB. %
 {To provide robustness against the noise for the analysis methods based on Proposition~\ref{prop:LPV_diss_polytopic} and~\ref{prop:LPV_diss_ellipsoidal}, we have introduced a regularization parameter in LMI~\eqref{eq:prop_LPV_diss_polytopic:a} to force the left-hand side to be less than $-\delta I$, and in LMI~\eqref{eq:prop_LPV_diss_ellipsoidal} to be larger than $\delta I$. The parameter~$\delta$ can be chosen as the standard deviation $\sigma_\mathrm{e}$ of the noise, which can be estimated in practice. While such a regularization introduces conservatism in the obtained performance bounds, it provides robust feasibility of the analysis against the noise.} The analysis results {under} the noisy data-dictionaries, where we took $\delta=\sigma_\mathrm{e}$,  are shown in Fig.~\ref{fig:influencenoise}.
The horizontal lines indicate the results for noise-free data, i.e., the earlier obtained results, while the blue circles and red crosses correspond to the upper bound $\gamma$ on the $(L-\tau)$-horizon $\ell_2$-gain calculated with Proposition~\ref{prop:LPV_diss_polytopic} and~\ref{prop:LPV_diss_ellipsoidal}, respectively. 
The plot shows that as {$\sigma_\mathrm{e}$ grows}, i.e., the SNR drops, the results of the analysis become more conservative. However, approaching to high noise levels in terms of a SNR $< 33$ dB, the optimization requires excessively large regularization to remain feasible, even beyond $\delta=\sigma_\mathrm{e}$ for Proposition~\ref{prop:LPV_diss_ellipsoidal}. This  indicates that our methods are applicable under low and moderate levels of noise (SNR $\geq 35$ dB), but a systematic stochastic way of handling the noise is required for further robustifying the analysis results against more severe noise conditions.

\subsection{Example II: {$(\mathit{L}-\tau)$-dissipativity for multiple $\mathit{L}$}}\label{ss:example2}
In this example, we analyze $(L-\tau)$-dissipativity for multiple values of $L$ and, with this, demonstrate the relationship between $(L-\tau)$-horizon dissipativity and infinite-horizon dissipativity in terms of the $\ell_2 $-gain. This example also illustrates the difference in computational load for the proposed methods. {Consider} a similar LPV system {as in Example \ref{example:1}}, now with $\dna=\dnb=2$. {In this case, the coefficients $a$ and $b$} are %
\begin{align*}
    \begin{bmatrix} a_{1,0} & a_{2,0}  \\ a_{1,1} & a_{2,1}  \end{bmatrix} & =%
    \begin{bmatrix} -0.00569  &  0.0706   \\ 0.1137    & -0.0210   \end{bmatrix} \\
    \begin{bmatrix} b_{1,0} & b_{2,0}  \\ b_{1,1} & b_{2,1}  \end{bmatrix} & =%
    \begin{bmatrix}   1.3735  & -0.2941   \\ 0.1254    &  0.6617   \end{bmatrix}.
\end{align*}
The scheduling set {$\mb{P}$} is defined as $[-0.2,\>0.2]$. We also impose a rate bound on the scheduling signal: %
$p_{k}-p_{k-1}\in[-0.1,\> 0.1]={\mb{D}}$. In this example, we determine upper bounds for the $(L-\tau)$-horizon $\ell_2 $-gain of this system using our direct dissipativity analysis approach{es} for $L=3,\dots,8$, and compare these results to the upper bound for the infinite-horizon $\ell_2 $-gain, which is obtained {via model-based analysis} using the \lpvcore toolbox.
\begin{figure}
\centering
\includegraphics[scale=1]{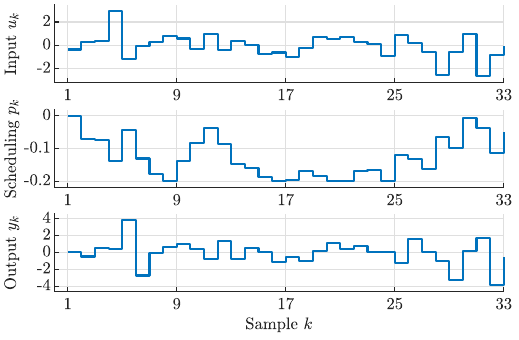}  %
\caption{Data-dictionary for Example II.}\label{fig:datadictionary2}  %
\end{figure}
We choose $\tau=2$ and generate our data-dictionary $\dataset$ with $N=33$, by exciting the system with an input  {$u_k\sim\mc{N}(0,1)$} and scheduling $p_k$, {whose samples are drawn from an i.i.d. uniform distribution {on $\mathbb{P}$} that is truncated to satisfy {the rate bound}. The resulting $\dataset$} is PE {of order $(L\le8,\dnx=2)$}. The generated data-dictionary is shown in Fig.~\ref{fig:datadictionary2}.

We use the model of the LPV system to compute {an upper bound $\gamma$ on the (infinite-horizon)} $\ell_2 $-gain with \lpvcore~{using a model-based approach}, which yields $\gamma=1.667$. {When~$\mb{D}$ is not considered in the analysis, we obtain $\gamma=1.754$}. With $\dataset$, we identify two LPV-SS representations using the same settings as in Example I, one considering $\mb{P}$ and $\mb{D}$ and one considering only $\mb{P}$. Calculating the upper bound $\gamma$ on the (infinite-horizon) $\ell_2 $-gain of the two identified models yields $\gamma=1.660$ for the former and $\gamma=1.754$ for the latter. 

We will now compare these (indirect) model-based analysis results to the results of direct data-driven $(L-\tau)$-dissipativity analysis for $L=3,\dots,8$. We will apply all four verification methods that we {have discussed} in this paper to this example, i.e., we verify \eqref{eq:thm_LPV_diss} by means of:
\begin{itemize}
	\item A polytopic description of $\mb{P}$, using the convex hull argument in Proposition~\ref{prop:LPV_diss_polytopic} ({denoted as} CHA),
	\item A quadratic description of $\mb{P}$, using the S-procedure in Proposition~\ref{prop:LPV_diss_ellipsoidal} ({denoted as} SP).
\end{itemize}
Furthermore, we reduce the conservatism {of the analysis} introduced in the aforementioned methods by:
\begin{itemize}
	\item Including {the rate bound $\mb{D}$ on $p$ in} the CHA method, as discussed in Section~\ref{sss:conservatism} ({denoted as} CHAr),
	\item Considering {$p$-}dependent multipliers and, as discussed in Remark~\ref{rem:PVmulti}, {solve SP with a sampling-based approach}  ({denoted as} SDM). {For this computation, we generate} $N_\mr{s}=1000$ sample trajectories from $\ms{P}_{[1,L]}$. 
\end{itemize}
{For the CHA(r) methods, we will use a constant $X$.}

\begin{figure}
\centering
\begin{subfigure}[b]{\linewidth}
\centering
\includegraphics[scale=1]{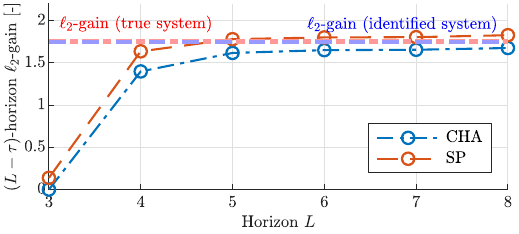}
\caption{}\label{fig:result_multL:con}
\end{subfigure}
\begin{subfigure}[b]{\linewidth}
\includegraphics[scale=1]{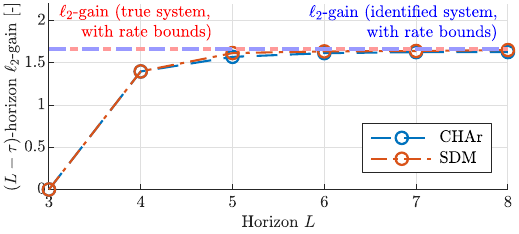}
\caption{}\label{fig:result_multL:ncon}
\end{subfigure}
\caption{Model-based and indirect data-driven dissipativity analysis versus direct data-driven $(L-\tau)$-dissipativity analysis of an LPV system for an increasing horizon  $L$. Plot (a) shows the results {when} the rate bounds on $p$ are not included in the analysis, while (b) shows the results {when} the {rate bounds are included.}}\label{fig:result_multL} %
\end{figure}

With the above {given} four methods, computing {an} upper bound $\gamma$ on the $(L-\tau)$-horizon $\ell_2 $-gain of the system using $\dataset$ {gives the results depicted in  Fig.~\ref{fig:result_multL} for $L=3,\dots,8$}. 
{The plot in} Fig.~\ref{fig:result_multL:con} shows {the results of the CHA and SP {data-driven} methods} %
together with the model-based and indirect dissipativity analysis methods, while the %
plot in Fig.~\ref{fig:result_multL:ncon} depicts the results of the {CHAr and SDM direct data-driven dissipativity analysis methods}
together with the model-based and indirect dissipativity analysis methods. %
The plots show for this simple system {that} the finite-horizon direct data-driven dissipativity analysis results converge relatively quick{ly} to the infinite-horizon results {of the} model-based approaches. As expected, Fig.~\ref{fig:result_multL:con} demonstrates that the dissipativity verification via Proposition~\ref{prop:LPV_diss_ellipsoidal} is slightly more conservative, i.e., the computed upper bound on the $\ell_2$-gain is slightly larger. However, this conservatism is negated when the multiplier is considered to be scheduling dependent, as shown in Fig.~\ref{fig:result_multL:ncon}. Hence, our methods give comparable results to the model-based approach when the horizon is chosen sufficiently large {even if} we \emph{only} make use of the information that is encoded in the data-dictionary $\dataset$. 

As highlighted throughout the paper, e.g., Remark~\ref{rem:manylmis}~and~\ref{rem:PVmulti}, the computational complexity of the CHA(r) method(s) can be significant, especially when compared to the method based on the S-procedure or the sampling-based method. This is {illustrated by} Table~\ref{tab:comp}, where we have listed the computation time for every method for an increasing $L$. For comparison, we also included the computation time for model-based $L$-dissipativity verification for the increasing horizon.
\begin{table}
\centering
\caption{Computation time (in seconds) for verifying $(L-\tau)$-dissipativity for $L=3,\dots,8$ using the methods discussed in this paper{:} Convex Hull Argument (CHA), S-Procedure (SP), CHA that incorporates rate bounds (CHAr) and sampling-based approach with Scheduling Dependent Multipliers (SDM). {The last column contains the model-based $(L-\tau)$-dissipativity verification approach (MBA) for comparison.}}\label{tab:comp}
\begin{tabular}{c|rrrrr}
$L$ & \multicolumn{1}{c}{CHA} & \multicolumn{1}{c}{SP} & \multicolumn{1}{c}{CHAr} & \multicolumn{1}{c}{SDM} & \multicolumn{1}{c}{MBA}  \\ \hline
3 & 0.82 & 0.51 & 1.10 & 9.65 & 0.41\\
4 & 1.85 & 0.38 & 3.99 & 15.00 & 0.32\\
5 & 4.31 & 0.39 & 9.10 & 12.31 & 0.33\\
6 & 6.90 & 0.40 & 21.39 & 10.88 & 0.35\\
7 & 16.55 & 0.42 & 56.25 & 12.91 & 0.32\\
8 & 38.31 & 0.45 & 168.80 & 12.08 & 0.36\\
\end{tabular} 
\end{table}
Due to the exponential growth of the number of constraints and decision variables {with horizon}  $L$ of the dissipativity verification problem using the CHA(r) methods, the computation time significantly increases for relatively small horizons, even for simple systems. This makes the approach based on the S-procedure much more suitable for larger systems/horizons at the price of conservatism.

{Finally, we want to comment on the choice of $L$. {We} have seen in this example that a larger $L$ yields {$\ell_2$-gains} `closer' to the infinite-horizon {$\ell_2 $-gain}, which is in line with the discussion of Remark~\ref{rem:fin_vs_inf}. 
In this particular case, we observed that the condition $(L-\tau)\ge2\dnr$ serves as a good heuristic for obtaining an $\ell_2$-gain estimate close to the true infinite-horizon value.
However, a generic guideline for choosing $L$ will require proper understanding {of} the fundamental relationship between $L$-dissipativity and infinite-horizon dissipativity, which is outside  the scope of this paper.}

\subsection{Example III: Unbalanced disc}
{As highlighted in Section~\ref{sec:introduction}, the LPV framework is often used for the analysis and control of nonlinear systems. Therefore, we will %
 {apply the proposed} data-driven dissipativity analysis methods {in a simulation study of} %
an unbalanced disc system, {showing} %
applicability of the methods on physical systems. In this example, we analyze $L$-dissipativity of the {unbalanced disc} 
in closed-loop with an LPV controller.

\subsubsection{Unbalanced disc system} 
The unbalanced disc system consists of a disc with an off-centered mass whose angular position can be controlled by {an} attached DC motor. {The system can be seen as a rotational} %
pendulum, {whose} continuous-time dynamics %
are described by
\begin{equation}\label{eq:unbalanced-disc}
    \ddot{\theta}(t)=-\tfrac{mgl}{J}\sin(\theta(t))-\tfrac{1}{\kappa}\dot{\theta}(t)+\tfrac{K_\mr{m}}{\kappa}u(t),
\end{equation}
where $\theta$ is the angular position of the disc in radians, $u$ is the input voltage to the system, which {can be used as a} control input, and $m,g,l,J,\kappa,K_\mr{m}$ are the physical parameters of the system, {given in} \cite{BoefCoxToth2021}. As we work in discrete time, we discretize the dynamics using a first-order Euler method with sampling-time $T_\mr{s}$, which yields
\begin{multline}\label{eq:UB_IO_NL}
    \theta_k + (\tfrac{T_\mr{s}}{\kappa}-2)\theta_{k-1} + (1 -\tfrac{T_\mr{s}}{\kappa}) \theta_{k-2} \\
    + T_\mr{s}^2\tfrac{mgl}{J}\sin(\theta_{k-2}) = T_\mr{s}^2\tfrac{K_\mr{m}}{\kappa}u_{k-2}. 
\end{multline}
We can \emph{embed} \eqref{eq:UB_IO_NL} {into an} LPV system by defining the scheduling as $p_k = \sinc(\theta_k)=\tfrac{\sin(\theta_k)}{\theta_k}$, which yields that $\mb{P}:=[-0.22,1]$. We choose $T_\mr{s}=0.01$ [s], which gives a negligible discretization error. %

\subsubsection{Controller design}
We design a reference tracking LPV controller for the unbalanced disc system according to the block-diagram depicted in Fig.~\ref{fig:controlscheme},
\begin{figure}
\centering
\includegraphics[width=\linewidth]{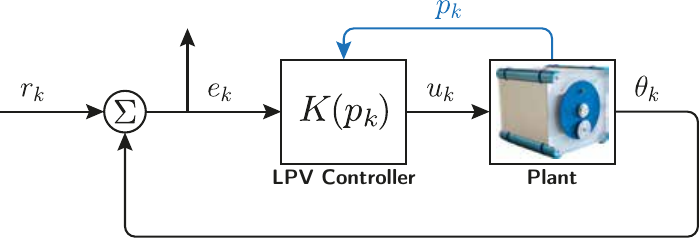}
\caption{Control scheme {considered in Example III}.}\label{fig:controlscheme}
\end{figure}
such that we can analyze $(L-\tau)$-dissipativity of the mapping $r_k\to e_k$, where $r_k$ represents the reference for $\theta$, and $e_k=r_k-\theta_k$. To use our developed methods, the closed-loop, i.e., the mapping $r_k\to e_k$, must admit a shifted-affine form as in \eqref{eq:sys}. This is ensured by designing  {a simple FIR-type of} LPV controller with shifted-affine scheduling dependence, i.e., 
\[
    u_k = k_0(p_k)e_k + k_1(p_{k-1})e_{k-1} + \dots + k_{n_\mr{k}}(p_{k-n_\mr{k}})e_{k-n_\mr{k}}.
\]
For $n_\mr{k}=2$, we design the LPV controller by manually tuning the coefficients $k_{0,0},\dots,k_{2,1}$. Choosing the coefficients as
\[
    \begin{bmatrix} k_{1,0} & k_{2,0} & k_{3,0} \\ k_{1,1} & k_{2,1} & k_{3,1} \end{bmatrix} = \begin{bmatrix}
        -80 & 80 & -3 \\ 1.5 & -3 & 0.3
    \end{bmatrix},
\]
yields a response as in Fig.~\ref{fig:closedloopresponse}.  
\begin{figure}
\centering
\includegraphics[scale=1]{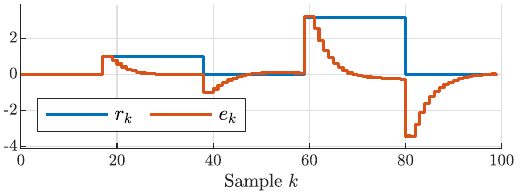}
\caption{Closed-loop response {of the unbalanced disc system with the considered LPV controller, which is used as the data-dictionary in Example III.}}
\label{fig:closedloopresponse}
\end{figure}
The closed-loop behavior can hence be described as a 4\tss{th}-order LPV-IO difference equation with shifted-affine scheduling dependence, where $r_k$ is the input, $e_k$ is the output and $p_k:=\sinc(r_k-e_k)$ is the scheduling.
\subsubsection{Closed-loop analysis}
As in the previous examples, we first calculate the upper bound $\gamma$ on the (infinite-horizon) $\ell_2$-gain of the closed-loop {with \lpvcore} system using the exact model, which yields $\gamma=1.44$. Note that, as $\mf{B}_\mr{NL}\subseteq\mf{B}_\mr{LPV}$, this implies {an} upper bound on the $\ell_2 $-gain of the nonlinear closed-loop system. For the data-driven dissipativity analysis, we calculate the finite-horizon $\ell_2$-gain for twice the order of the system, i.e., $(L-\tau)=8$ with $L=12$ and $\tau=4$. Furthermore, we construct our data-dictionary $\dataset$ directly from the response depicted in Fig.~\ref{fig:closedloopresponse}, which {corresponds} to taking nominal operational data from the system, instead of {a} carefully designed experiment. Note that for this case, the scheduling data in $\dataset$ is constructed via $p_k=\sinc(r_k-e_k)$. We determine {the} $(L-\tau)$-dissipativity {based $\ell_2 $-gain} with the SP, SDM and the MBA approaches. For the SDM approach, we sample 100 trajectories from $\ms{P}_{[1,L]}$. The analysis results with {all the considered} approaches are given in Table~\ref{tab:unbal}.
\begin{table}
\centering
\caption{$(L-\tau)$-dissipativity analysis on the LPV embedding of the controlled unbalanced disc system for $L=12$, $\tau=4$ using the SP, SDM and MBA approaches.}\label{tab:unbal}
\begin{tabular}{c|cccc}
 & {True $\ell_2 $-gain}& SP & SDM & MBA  \\ \hline
Upper bound $\gamma$ & 1.449 & 1.548 & 1.319 & 1.263 \\
\end{tabular}
\end{table}
We {can} see that the upper bound given by \lpvcore is between the obtained bounds of the more conservative SP and the SDM {approaches}. (Note that the MBA and the SDM approaches under-estimate the infinite-horizon $\ell_2 $-gain, which is due to the difference between finite- and infinite-horizon dissipativity, see also Remark~\ref{rem:fin_vs_inf}). When we increase $L$ to 25, the resulting bound on the finite-horizon $\ell_2 $-gain with the SDM approach is 1.325, while the SP gives 1.807. We can conclude that application of the presented methods on LPV embeddings of nonlinear systems can yield, for sufficiently large $L$, good indications of (an upper bound on) the infinite-horizon $\ell_2 $-gain. %
}

\section{Conclusions and future work}
\label{sec:conclusions}
This paper proposes a direct data-driven dissipativity analysis method for LPV systems
 that can be represented in an IO form with shifted-affine dependence on the scheduling signal. Using a data-driven representation of the LPV system, which is constructed from a data set with persistently exciting inputs and scheduling,  we can analyze {its} finite-horizon dissipativity property %
 for any quadratic performance specification {by our proposed methods}. This allows to give performance guarantees {for} the considered system using only a single data set measured from the system. As we show in the paper, the analysis can be accomplished by solving an SDP subject to LMI constraints, which is constructed from the data set and {the considered restrictions} %
  of the scheduling set. By means of the presented examples, we {have} demonstrated that our methods {give similar results as %
  the classical methods, which require the availability of a model of the system.} %
{We also show} that the finite-horizon dissipativity property converges to the infinite-horizon dissipativity property (employed in the model-based methods) already for relatively short horizons. 

For future work, we aim to analyze the convergence to the infinite-horizon dissipativity property, extend our results to the dissipativity formulation of Willems \cite{Willems1972} that includes a storage functions and investigate the handling of noisy data.

\appendices
\section{Model-based $L$-dissipativity}\label{app:modelbasedLdissip}
In model-based dissipativity analysis, {definition and testing of} %
classical dissipativity, see \cite{Willems1972, HillMoylan1980}, {has been throughly investigated. However, $L$-dissipativity has not been directly considered.} 
{In this section, we give a model-based equivalent of Theorem~\ref{thm:LPV_diss} in terms of an $L$-dissipativity test}.

{Note that LPV systems represented by \eqref{eq:sys} have a direct minimal state-space realization in the form of}
\begin{align*}
    x_{k+1} & = A(p_k) x_k + B(p_k)u_k, \\
    y_k & = C(p_k) x_k + D(p_k)u_k.
\end{align*}
{as discussed in \cite{Toth2011_SSrealizationTCST}. This form is easily computable in the SISO case, while computation in the {MIMO} case requires the choice of an algebraically independent state basis, which can be accomplished via \cite{Toth15CDCb}.} 

It is well-known that any sequence $y_{[1,L]}$ can be written as
\begin{equation}
    \vect{y}_L = \ms{O}_L(p_{[1,L]}) x_1 + \ms{T}_L(p_{[1,L]}) \vect{u}_L,
\end{equation}
with
\begin{equation}
    \ms{O}_L(p_{[1,L]}) = \begin{bmatrix} C(p_1) \\ C(p_2)A(p_1) \\ \vdots \\ C(p_L)A(p_{L-1})\cdots A(p_1) \end{bmatrix}
\end{equation}
and
\begin{multline}
    \ms{T}_L(p_{[1,L]}) = \\
    \begin{bsmallmatrix} D(p_1) & 0 & \cdots & 0 \\ C(p_2)B(p_1) & D(p_2) & \sddots & \svdots \\ C(p_3)A(p_2)B(p_1) & C(p_3)B(p_2) & \sddots & \svdots  \\ \svdots & \sddots & \sddots & \svdots \\ C(p_L)A(p_{L-1})\cdots A(p_2)B(p_1) & \cdots & \cdots & D(p_L) \end{bsmallmatrix}.
\end{multline}
By the assumption of zero initial conditions in the definition for $L$-dissipativity, we have that
\begin{equation}
    \vect{y}_L = \ms{T}_L(p_{[1,L]}) \vect{u}_L.
\end{equation}
Now considering \eqref{eq:L_diss_stacked}, we can write
\begin{equation}
    0 \leq \begin{bmatrix} {\vect{u}}_L\\ {\vect{y}}_L\end{bmatrix}^{\!\top} \! \Pi_L \begin{bmatrix} {\vect{u}}_L\\ {\vect{y}}_L \end{bmatrix} = {\vect{u}}_L^\top \begin{bmatrix} I_L \\ \ms{T}_L(p_{[1,L]})\end{bmatrix}^{\!\top} \! \Pi_L \begin{bmatrix} I_L\\ \ms{T}_L(p_{[1,L]}) \end{bmatrix} \vect{u}_L.
\end{equation}
Hence, model-based $L$-dissipativity verification corresponds to
\begin{equation}\label{eq:LMImodelbased}
    \begin{bmatrix} I_L \\ \ms{T}_L(p_{[1,L]})\end{bmatrix}^\top \Pi_L \begin{bmatrix} I_L\\ \ms{T}_L(p_{[1,L]}) \end{bmatrix} \possemidef0,
\end{equation}
for all scheduling trajectories in $\ms{P}_{[1,L]}$. Based on the form of $\ms{P}_{[1,L]}$, e.g., polytopic or quadratic, one can select an existing method to verify \eqref{eq:LMImodelbased} in a computationally attractive manner.

\bibliographystyle{IEEEtran}
\bibliography{refs_ddd}

 \begin{IEEEbiography}[{\includegraphics[width=1in,height=1.25in,clip,keepaspectratio]{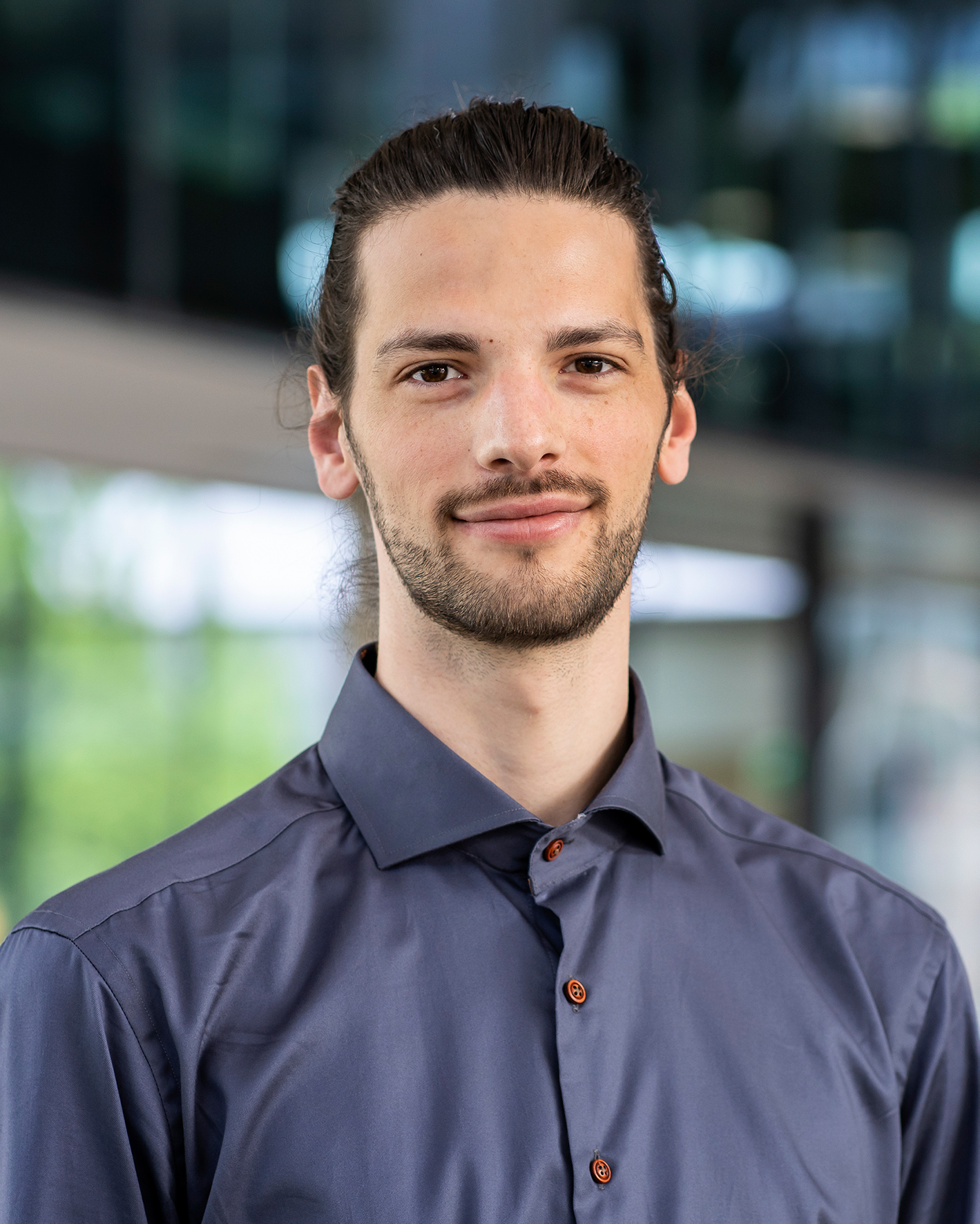}}]{Chris Verhoek} received his B.Sc. degree in Mechatronics from the Avans University of Applied Sciences and M.Sc. degree (Cum Laude) in Systems and Control from the Eindhoven University of Technology (TU/e), in 2017 and 2020 respectively. His M.Sc. thesis was selected as best thesis of the Electrical Engineering department in the year 2020. He is currently pursuing a Ph.D. degree under the supervision of Roland T\'oth and Sofie Haesaert at the Control Systems Group, Dept. of Electrical Engineering, TU/e. In the fall of 2023, he was a visiting researcher at the ETH Z{\"u}rich, Switzerland. His main research interests include (data-driven) analysis and control of nonlinear and LPV systems and learning-for-control techniques with stability and performance guarantees.
\end{IEEEbiography}

\begin{IEEEbiography}[{\includegraphics[width=1in,height=1.25in,clip,keepaspectratio]{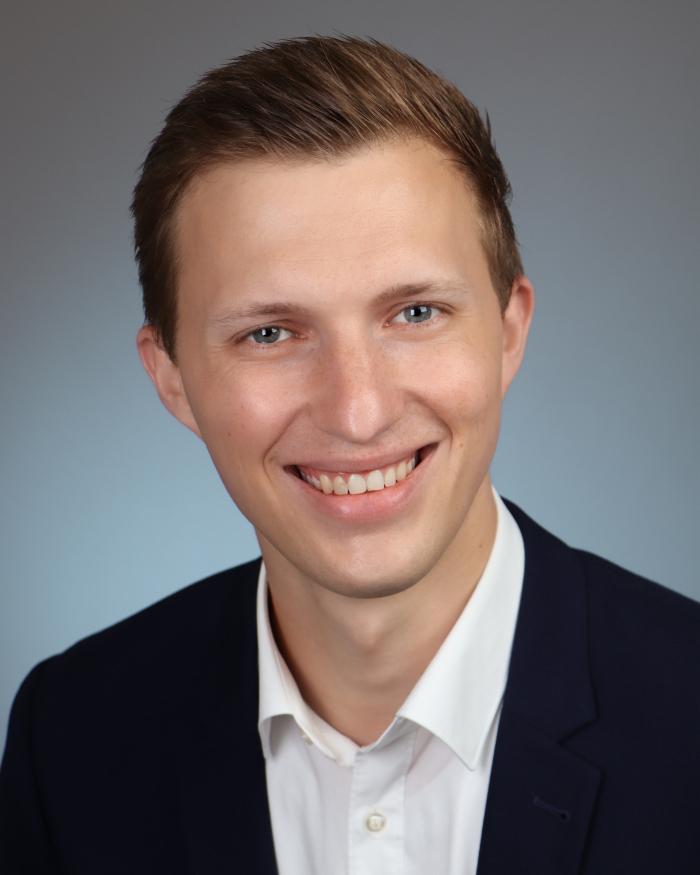}}]{Julian Berberich} received a Master's degree in Engineering Cybernetics from the University of Stuttgart, Germany, in 2018. In 2022, he obtained a Ph.D. in Mechanical Engineering, also from the University of Stuttgart, Germany. He is currently working as a Lecturer (Akademischer Rat) at the Institute for Systems Theory and Automatic Control at the University of Stuttgart, Germany. In 2022, he was a visiting researcher at the ETH Z{\"u}rich, Switzerland. He has received the Outstanding Student Paper Award at the 59th IEEE Conference on Decision and Control in 2020 and the 2022 George S. Axelby Outstanding Paper Award. His research interests include data-driven analysis and control as well as quantum computing.
\end{IEEEbiography}

\begin{IEEEbiography}[{\includegraphics[width=1in,height=1.25in,clip,keepaspectratio]{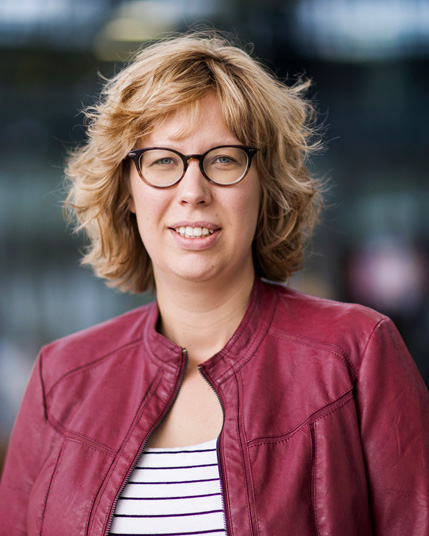}}]{Sofie Haesaert} received the B.Sc. degree Cum Laude in mechanical engineering and the M.Sc. degree Cum Laude in systems and control from the Delft University of Technology (TUDelft), Delft, The Netherlands, in 2010 and 2012, respectively, and the Ph.D. degree from TU/e, Eindhoven, The Netherlands, in 2017.

She is currently an Assistant Professor with the Control Systems Group, Department of Electrical Engineering, TU/e. From 2017 to 2018, she was a Postdoctoral Scholar with Caltech. Her research interests are in the identification, verification, and control of cyber-physical systems for temporal logic specifications and performance objectives.
\end{IEEEbiography}

\begin{IEEEbiography}[{\includegraphics[width=1in,height=1.25in,clip,keepaspectratio]{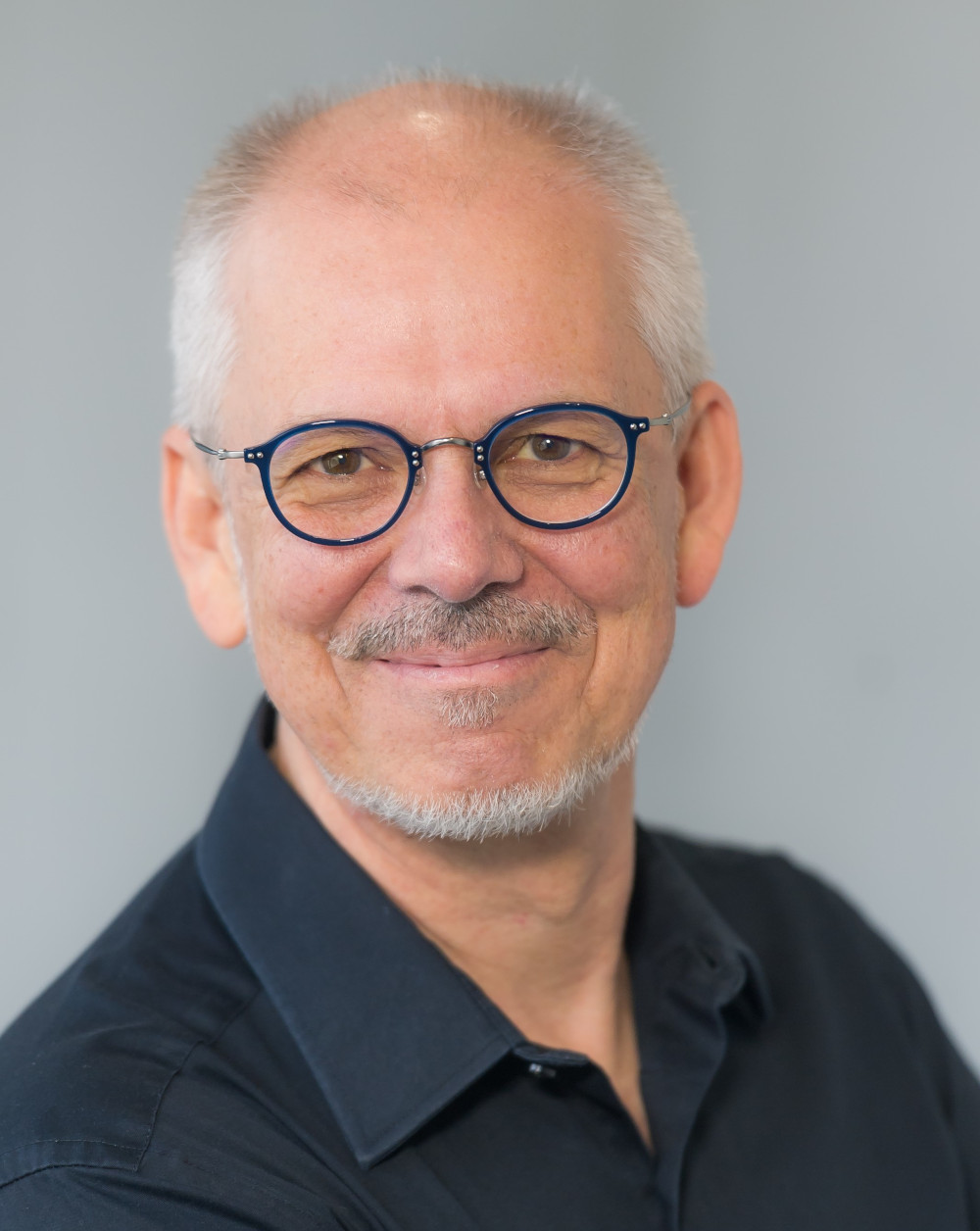}}]{Frank Allg\"ower} is professor of mechanical engineering at the University of Stuttgart, Germany, and Director of the Institute for Systems Theory and Automatic Control (IST) there.

Frank is active in serving the community in several roles: Among others he has been President of the International Federation of Automatic Control (IFAC) for the years 2017-2020, Vice-president for Technical Activities of the IEEE Control Systems Society for 2013/14, and Editor of the journal Automatica from 2001 until 2015. From 2012 until 2020 Frank served in addition as Vice-president for the German Research Foundation (DFG), which is Germany's most important research funding organization. 

His research interests include predictive control, data-based control, networked control, cooperative control, and nonlinear control with application to a wide range of fields including systems biology.
\end{IEEEbiography}

\begin{IEEEbiography}[{\includegraphics[width=1in,height=1.25in,clip,keepaspectratio]{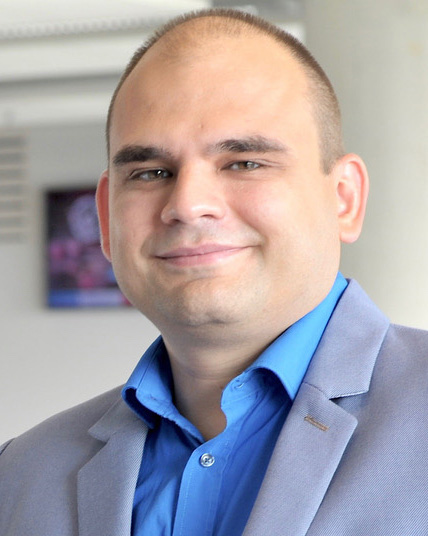}}]{Roland T\'oth} received his Ph.D. degree with Cum Laude distinction at the Delft Center for Systems and Control (DCSC), TUDelft, Delft, The Netherlands in 2008.  He was a Post-Doctoral Research Fellow at TUDelft in 2009 and Berkeley in 2010. He held a position at DCSC, TUDelft in 2011-12. Currently, he is a Full Professor at the Control Systems Group, TU/e and a senior researcher at HUN-REN Institute for Computer Science and Control (SZTAKI) in Budapest, Hungary. He is a Senior Editor of the IEEE Transactions on Control Systems Technology. 

His research interests are in identification and control of linear parameter-varying (LPV) and nonlinear systems, developing machine learning methods with performance and stability guarantees for modeling and control, model predictive control and behavioral system theory with a wide range of applications, including precision mechatronics and autonomous vehicles.
\end{IEEEbiography}

\end{document}